\documentclass[twocolumn,twocolappendix]{aastex631}
\usepackage{textcomp}

\usepackage{multirow}

\usepackage{wrapfig}

\usepackage{booktabs}
\usepackage{natbib}

\usepackage{verbatim}
\usepackage{graphicx}
\usepackage{color}
\usepackage{amssymb}
\usepackage{hhline}
\usepackage{amsmath}
\usepackage{hyperref}
\usepackage{cleveref}

\usepackage[english]{babel}
\usepackage[autostyle, english = american]{csquotes}

\def\micron{$\mu$m}

\def\nh3{NH$_{3}$}

\def\arcsec{$^{\prime\prime}$}
\def\arcmin{$^{\prime}$}

\newcommand{\Msol}{$M_{\odot}$}

\begin{document}

\title{3D-Herschel: Constraining Dust Emission with Panchromatic Modeling of 3D-HST Galaxies}
\shorttitle{Panchromatic SED Fitting of 3D-Herschel}
\shortauthors{McNulty et al. 2026}

\author[0009-0007-3915-2078]{Seamus McNulty}
\affiliation{Department of Astronomy, University of Massachusetts, Amherst, MA, 01003 USA}
\email{Seamusjmcn@gmail.com}

\author[0000-0002-8442-3128]{Mimi Song}
\affiliation{Department of Astronomy, University of Massachusetts, Amherst, MA, 01003 USA}

\author[0000-0001-7160-3632]{Katherine E. Whitaker}
\affiliation{Department of Astronomy, University of Massachusetts, Amherst, MA, 01003 USA}
\affiliation{Cosmic Dawn Center (DAWN), Niels Bohr Institute, University of
Copenhagen, Jagtvej 128, København N, DK-2200, Denmark}

\author[0000-0001-6755-1315]{Joel Leja}
\affiliation{Department of Astronomy \& Astrophysics, The Pennsylvania State University, University Park, PA 16802, USA}
\affiliation{Institute for Computational \& Data Sciences, The Pennsylvania State University, University Park, PA 16802, USA}
\affiliation{Institute for Gravitation and the Cosmos, The Pennsylvania State University, University Park, PA 16802, USA}

\author[0000-0002-8450-9992]{Aubrey Medrano}
\affiliation{Department of Astronomy, University of Massachusetts, Amherst, MA, 01003 USA}

\author[0000-0003-0384-0681]{Elijah P. Mathews}
\affiliation{Department of Astronomy \& Astrophysics, The Pennsylvania State University, University Park, PA 16802, USA}
\affiliation{Institute for Computational \& Data Sciences, The Pennsylvania State University, University Park, PA 16802, USA}
\affiliation{Institute for Gravitation and the Cosmos, The Pennsylvania State University, University Park, PA 16802, USA}

\author[0000-0001-5414-5131]{Mark Dickinson}
\affiliation{NSF’s National Optical-Infrared Astronomy Research Laboratory, 950 N. Cherry Avenue, Tucson, AZ 85719, USA}

\author[0000-0003-4268-0393]{Hanae Inami}
\affiliation{Hiroshima Astrophysical Science Center, Hiroshima University, 1-3-1 Kagamiyama, Higashi-Hiroshima, Hiroshima 739-8526, Japan}

\author[0000-0002-2057-5376]{Ivo Labbe}
\affiliation{Centre for Astrophysics and Supercomputing, Swinburne University of Technology, Melbourne, VIC 3122, Australia}

\author[0000-0001-9002-3502]{Danilo Marchesini}
\affiliation{Department of Physics and Astronomy, Tufts University, 574 Boston Ave., Medford, MA 02155, USA}

\author[0000-0001-8592-2706]{Alexandra Pope}
\affiliation{Department of Astronomy, University of Massachusetts, Amherst, MA, 01003 USA}

\author[0000-0003-4702-7561]{Irene Shivaei}
\affiliation{Centro de Astrobiología (CAB), CSIC-INTA, Carretera de Ajalvir km 4, Torrejón de Ardoz 28850, Madrid, Spain}
\affiliation{Steward Observatory, University of Arizona, Tucson, AZ 85721, USA}

\begin{abstract} 

We present 3D-Herschel, a new publicly released 0.3-350$\mu$m photometric catalog that combines deblended Herschel far-infrared (FIR) imaging with the CANDELS/3D-HST legacy fields to probe the dust-obscured universe.
Using the 17-parameter Bayesian fitting code {\fontfamily{qcr}\selectfont Prospector-$\beta$}, we model 41,387 galaxies spanning 0.5 $<$ \textit{z} $<$ 2.5 to measure stellar and dust properties. Comparing fits with and without FIR constraints, we find that for the 3.2$\%$ of galaxies with $>3 \sigma$ detections in $\geq2$ Herschel bands, UV-MIR-only models (0.3-24$\mu$m) recover robust stellar ages, SFRs, and stellar masses (50-70$\%$ within the median 1$\sigma$ error). Consequently, the {Prospector-$\beta$} star-forming sequence is unchanged by the inclusion of FIR data (average deviation 0.1$\pm$0.07 dex between UV-MIR and UV-FIR fits at fixed stellar mass), confirming that the offset relative to UV+IR-based estimates reported by \citet{Leja2022} is robust to the lack of direct FIR constraints. However, the use of rigid log-average IR templates with fixed dust emission parameters ($\gamma$, $U_{\mathrm{min}}$, $q_{\mathrm{PAH}}$) in UV-MIR modeling yields cold dust temperatures ($\sim$7K colder than Herschel-informed fits at all redshifts) and an unevolving MIR-to-IR luminosity ratio, with $\sim$0.2 dex lower IR-to-7.7$\mu$m luminosity ratios at the low-mass end of Herschel-detected galaxies (log($M_{\star}$) $\sim$ 9.6 $M_{\odot}$). These results demonstrate that MIR-to-IR conversions depend on stellar mass, cautioning against $L_{\mathrm{IR}}$-independent templates without FIR data. For galaxies with $<10^{11} \ M_{\odot}$ at $z>1.5$, Herschel can at best provide upper limits due to source confusion; next-generation FIR telescopes will be essential for distant galaxies.

\end{abstract}


\section{Introduction} \label{sec:intro}

Studying galaxy properties during cosmic noon ($z\sim1-3$), the peak of both cosmic star formation \citep{Madau2014} and dust-obscured activity \citep{Casey2018, Casey2021}, provides crucial insights into the physical processes that drive their growth and evolution \citep[e.g.,][]{Behroozi2013, vanDokkum2015}. Because direct measurements of galaxy properties are not possible, astronomers rely on indirect methods, inferring stellar populations, dust, and gas from observed light. Much of this radiation is heavily attenuated by dust \citep{Madau2014, Casey2018}, especially among the most massive galaxies \citep{Whitaker2017}, with the complex interplay among different galaxy properties further complicating interpretation \citep[e.g.,][]{Leja2019a}. To disentangle these effects, models are fit to photometric data to reconstruct a galaxy's spectral energy distribution (SED) \citep{JohnsonLejaConroy2021, Carnall2018, Iyer2019}. Constraining these models requires broad photometric coverage, ideally spanning the rest-ultraviolet to far-infrared (UV–FIR), where both stellar and dust emission are captured \citep{Conroy2013}.

Some of the deepest FIR windows into the universe (70–350$\mu$m) lie within the Cosmic Assembly Near-infrared Deep Extragalactic Legacy Survey (CANDELS)/3D-HST extragalactic fields \citep{Grogin2011, Koekemoer2011, Skelton2014, Momcheva2016}, regions extensively covered by Spitzer Space Telescope and Herschel Space Observatory. While public FIR-selected catalogs exist for almost all Herschel surveys in these fields \citep[e.g.,][]{Elbaz2011, Oliver2012, Magnelli2013, Jin2018, Liu2018, Shirley2021}, simple cross-matching with ultraviolet to near-infrared (NIR) catalogs will not yield robust photometry due to severe blending issues caused by the low-resolution nature of the data. Instead, ``forced photometry'', which uses source positions and light profiles from a higher resolution prior, identifies roughly 50\% more FIR-bright galaxies and enables an analysis reaching fainter fluxes than is otherwise possible \citep[e.g.,][]{Lang2016}.

Combining Herschel’s long wavelength coverage through forced photometry with the existing high-quality 3D-HST photometric catalogs \citep{Skelton2014}, yields a dataset of broad- and medium-band photometry plus grism spectroscopy \citep{Momcheva2016}. A UV–FIR photometric catalog enables stellar population synthesis (SPS) modeling of galaxy SEDs observed during the most dust-obscured, star-forming epoch of the universe.

Modern SPS studies have moved beyond the four-to-six parameter models that left substantial systematic uncertainties \citep{BrinchmannEllis2000, Papovich2001, Shapley2001, Salim2007, Kriek2009, Maraston2010, Acquaviva2011} toward high-dimensional (N$\gtrsim$7) Bayesian frameworks \citep{HanHan2014, Leja2019a, Boquien2019, Bowman2020}. {\fontfamily{qcr}\selectfont Prospector-$\beta$} \citep{Wang2023} is a 17-parameter example that combines a nonparametric star formation history (SFH), a two-component dust attenuation model based on \citet{CharlotFall2000}, and dust emission powered via energy balance \citep{daCunha2008} with free dust emission parameters. The nonparametric SFH avoids the prior-driven biases in stellar mass, SFR, and age incurred when fixed functional forms are imposed \citep[$\gtrsim$0.1, 0.3, 0.2 dex respectively;][]{Carnall2019}. The two-component dust model separates birth-cloud attenuation around young stars from diffuse ISM attenuation \citep{Leja2017, Conroy2013}, mitigating the stellar mass underestimates experienced when a universal attenuation curve is assumed \citep{KriekConroy2013}. Finally, allowing polycyclic aromatic hydrocarbon (PAH) strengths to vary \citep{DraineLi2007, Leja2017, Leja2019b, Abdurro2021} enables more realistic estimates of dust re-emission. Together these flexibilities help break degeneracies among dust, age, and mass \citep{Leja2019a}.

These modeling improvements translate into systematically different SFR estimates than those inferred from combined rest-UV and IR fluxes \citep{Daddi2007, KennicuttEvans2012, Whitaker2012, Whitaker2014}. UV+IR estimates remain broadly consistent for actively star-forming systems but can overestimate SFRs in galaxies with reduced star formation activity by up to $\sim$1 dex \citep{Utomo2014, Hayward2015, Fumagalli2014}, because some IR luminosity is reprocessed light from older ($>$100 Myr) stars rather than current star formation \citep{Leja2019b}. By disentangling these contributions, {\fontfamily{qcr}\selectfont Prospector} yields SFRs $\sim$0.1-1 dex lower in passive systems and a star-forming main sequence (SFMS) lying $\sim$0.2-0.5 dex below traditional UV+IR estimates \citep{Leja2022}, implying an older and more massive universe. However, recent studies employing these improved modeling methods remain limited to UV-MIR coverage \citep{Leja2019b, Leja2022, Mathews2023, Shivaei2024}, with the reddest filters reaching only 24\micron\ and lacking constraints on the IR SED past this. 

At the time of Herschel's launch, the possibility of directly constraining $L_{\mathrm{IR}}$, and thus incorporating FIR emission into full SED modeling, represented a major advance in quantifying dust emission \citep[e.g.,][]{Burgarella2005, Noll2009} and breaking long-standing degeneracies. Early Herschel results showed tight correlations between MIR and total IR luminosity (e.g., via the IR8 relation; \citealt{Elbaz2011, Reddy2012}), which motivated the construction of fixed IR templates \citep[e.g.,][]{Whitaker2014}. Subsequent work demonstrated that these templates must vary with physical conditions such as the PAH mass fraction, stellar mass, and redshift \citep{Shivaei2017, Whitaker2017}, yet how much fixed-template assumptions bias inferred dust temperatures and MIR-to-IR ratios in modern Bayesian SPS frameworks has not been systematically quantified. Ultimately, however, Herschel's PACS/SPIRE data are strongly confusion-limited in deep extragalactic fields, preventing robust flux measurements for most individual galaxies and forcing reliance on templates or stacking. Overcoming this limitation requires improved positional and photometric priors to extract reliable fluxes below the confusion limit---an approach we adopt here by deblending the Herschel images to recover significantly higher FIR photometry.


In this paper, we present a comprehensive, long-wavelength analysis of how FIR data influence galaxy scaling relations using {\fontfamily{qcr}\selectfont Prospector} modeling. We deblend low-resolution Herschel images to incorporate FIR photometry into the 3D-HST photometric catalogs \citep{Skelton2014, Momcheva2016}, producing a new UV-FIR (0.3-350$\mu$m) dataset, 3D-Herschel. By jointly modeling UV-NIR starlight and dust re-emission in the MIR-FIR, we quantify the impact of adding FIR constraints to SPS modeling. Coupled with {\fontfamily{qcr}\selectfont Prospector}'s 17-parameter Bayesian framework, this approach improves constraints on dust emission and helps to mitigate well-known degeneracies in SPS modeling.

The paper is organized as follows. In \textsection \ref{sec:data}, we describe the data used in this study, including the 3D-HST catalogs, Herschel observations, and the forced photometry methods employed to extract deblended FIR fluxes. \textsection \ref{sec:validation} presents the validation of the new 3D-Herschel catalog. In \textsection \ref{sec:modeling} we present a comparison between {\fontfamily{qcr}\selectfont Prospector} fits for 3D-HST, and a more efficient emulator, adopted for the 3D-Herschel analysis. Results from the SED modeling are given in \textsection \ref{sec:results}, with a higher-quality, IR-bright subsample analyzed in \textsection \ref{sec:IR bright}, and the conclusion is presented in \textsection \ref{sec:conclusion}.


Throughout the paper, we adopt a concordance $\Lambda$CDM cosmology with $H_0 = 70$ km s$^{-1}$ Mpc$^{-1}$, $\Omega_M = 0.3$, and $\Omega_{\Lambda} = 0.7$.
We use a \citet{Chabrier2003} initial mass function (IMF) between 0.1 \Msol\, and 100 \Msol. All uncertainties are quoted at the 68\% confidence level.


\begin{deluxetable*}{lccccccccc}
\tablecaption{Summary of Herschel programs\label{tab:herscheldata}}
\tablehead{
\colhead{Field} & \colhead{Area (arcmin$^2$)\tablenotemark{e}} &
\multicolumn{2}{c}{Survey} & \colhead{70$\mu$m} & \colhead{100$\mu$m} &
\colhead{160$\mu$m} & \colhead{250$\mu$m} & \colhead{350$\mu$m} \\
\cline{3-4} \cline{5-9}
& & \colhead{PACS} & \colhead{SPIRE} & \multicolumn{5}{c}{Depth (mJy, 1$\sigma$)}
}
\startdata
GOODS-S & 177 & GOODS-H, PEP & HerMES & 0.3\tablenotemark{a} & 0.2\tablenotemark{a} & 0.4\tablenotemark{a} & 0.9\tablenotemark{d} & 0.8\tablenotemark{d} \\
GOODS-N & 164 & GOODS-H, PEP & GOODS-H & -- & 0.3\tablenotemark{a} & 0.7\tablenotemark{a} & 1.9\tablenotemark{c} & 2.4\tablenotemark{c} \\
COSMOS  & 199 & CANDELS-H, PEP & CANDELS-H & -- & 0.5\tablenotemark{b} & 1.0\tablenotemark{b} & 1.6\tablenotemark{f} & 1.3\tablenotemark{f} \\
UDS     & 201 & CANDELS-H & CANDELS-H & -- & 2.2\tablenotemark{g} & 3.4\tablenotemark{b} & 2.2\tablenotemark{f} & 1.9\tablenotemark{f} \\
\enddata
\vspace{-1mm}
\tablenotetext{a}{\scriptsize Depths for GOODS-S and GOODS-N bands are taken from the PEP public data release documentation.}
\vspace{-1mm}
\tablenotetext{b}{\scriptsize Depths for UDS and COSMOS PACS bands are taken from \citet{Popesso2019}.}
\vspace{-1mm}
\tablenotetext{c}{\scriptsize From \citet{Elbaz2011}.}
\vspace{-1mm}
\tablenotetext{d}{\scriptsize From \citet{Oliver2012}.}
\vspace{-1mm}
\tablenotetext{e}{\scriptsize From \citet{Skelton2014}.}
\vspace{-1mm}
\tablenotetext{f}{\scriptsize From \url{http://www.astrodeep.eu/data/}.}
\vspace{-1mm}
\tablenotetext{g}{\scriptsize Estimated based on our flux comparisons (Figure \ref{fig:flux comparisons}), as the original source is direct detection from Spitzer, opposed to the forced photometry methods used in this study.}

\tablecomments{Quoted depths are median 1$\sigma$ flux uncertainties for each field. For GOODS-S and GOODS-N, we list the depth in the deepest region.}
\end{deluxetable*}

\section{Data} \label{sec:data}

\subsection{3D-HST/CANDELS}

The parent sample for this study originates from the public multi-wavelength 3D-HST photometric catalog, covering 0.3-8$\mu$m \citep{Skelton2014} \footnote{\url{http://3dhst.research.yale.edu}}. The 3D-HST survey leverages imaging from the CANDELS program, designed to study galaxy evolution \citep{Skelton2014, Momcheva2016}.
The original sample consists of $\sim$166k sources detected in HST \textit{J$_{125}$+JH$_{140}$+H$_{160}$} images covering $\sim$740 arcmin$^2$ over the four deep CANDELS fields, COSMOS, GOODS-S, GOODS-N, UDS. Beyond this HST imaging, observations additionally include a suite of ground-based imaging, Spitzer 4-8\micron\ IRAC forced photometry \citep{Skelton2014}, with a supplement of 24\micron\ Spitzer/MIPS forced photometry supplied by \citet{Whitaker2014}. There are up to 45 total photometric bands in the original catalogs.

The Spitzer/MIPS 24$\mu$m photometry presented by \citet{Whitaker2014} was derived using a forced-photometry approach closely analogous to that adopted for the Herschel imaging in this work. High resolution HST/WFC3 imaging was used as a positional and morphological prior, enabling fluxes to be measured at fixed source positions while modeling and deblending contributions from neighboring objects in the low-resolution MIPS data. The HST detection images were PSF-matched to the F160W resolution and rebinned to account for the large MIPS point-spread function (FWHM $\approx$ 6", comparable to that of Herschel/PACS at 100$\mu$m). Photometry was performed using the Multi-resolution Object PHotometry oN Galaxy Observations ({\fontfamily{qcr}\selectfont MOPHONGO}; \citealt{Labbe2006}) code, which simultaneously fits all sources within each beam using position-dependent convolution kernels, thereby mitigating source confusion and blending. This methodology is directly relevant to the challenges inherent to Herschel photometry, where similarly large beam sizes necessitate the use of high-resolution priors for robust flux extraction.

Slitless spectroscopy covers $\sim$75\% of the CANDELS area, providing redshift estimates with precisions of order $\frac{\Delta z}{1+z} \sim 0.003$ when combined with broad and medium-band photometry \citep{Skelton2014, Momcheva2016}. We adopt redshifts, $z_{\rm best}$, from \citet{Momcheva2016}, which are ranked in order depending on their availability. The 'best' starting with ground-based spectroscopic redshifts, next space-based grism redshifts provided by the 3D-HST grism survey \citep{Momcheva2016}, and finally otherwise photometric redshifts estimated from the EAZY photometric redshift code \citep{Brammer2008}. 
We refer the reader to \citet{Skelton2014} and \citet{Momcheva2016} for a full description of the 3D-HST data and catalog construction. 

\subsection{Herschel data}

We use the deepest available Herschel imaging obtained with PACS (70--160$\mu$m) and SPIRE (250--350$\mu$m) in four CANDELS fields. The key Herschel programs used in this study are the Great Observatories Origins Deep Survey-Herschel (GOODS-Herschel) (PI D. Elbaz; \citealt{Elbaz2011}) and CANDELS-Herschel (PI M. Dickinson).
These two surveys are the deepest Herschel observations to extend the capabilities of CANDELS/3D-HST by incorporating FIR observations, aiming to probe typical star-forming galaxies up to $z\sim2$.

In brief, the GOODS-Herschel survey obtained both PACS and SPIRE imaging in the GOODS-N region, contributing to a comprehensive dataset for this field. In GOODS-S, the survey focused on obtaining ultradeep PACS imaging in a sub-region of 10\arcmin $\times$ 10\arcmin.
Likewise, the CANDELS-Herschel covers the COSMOS and UDS fields, both in PACS and SPIRE, to a comparable depth to the GOODS fields.

We also utilize Herschel data from the PACS Evolutionary Probe (PEP) and the Herschel Multi-tiered Extragalactic Survey (HerMES).
PEP is a survey conducted with PACS, spanning over 2.7 deg$^2$ targeting well-known extragalactic fields, including the CANDELS fields \citep{Lutz2011}, while
HerMES covers a wide area of 70 deg$^2$, with both the PACS and SPIRE instruments \citep{Oliver2012}.

The Herschel PACS and SPIRE maps used in this work for GOODS-N, COSMOS, and UDS, as well as PACS-only maps for GOODS-S, were constructed by H. Inami (private communication) by combining observations from the GOODS-H, CANDELS-H, and PEP surveys. For GOODS-S SPIRE data, we instead use the public HerMES maps. Table \ref{tab:herscheldata} lists the area, programs, and $1\sigma$ depths for the Herschel images used in this study. 

We do not include the AEGIS field in the present analysis because the corresponding photometric catalog could not be independently validated to the same standard as the other 3D-HST fields. Comparisons with external Herschel-based catalogs show large systematic discrepancies in the SPIRE fluxes ($\sim$2 magnitudes), indicating unresolved calibration or reduction issues. Resolving these discrepancies would require a full reprocessing of the AEGIS Herschel data, which is beyond the scope of this paper. 

\subsubsection{Forced photometry} \label{subsec:ForcedPhotometry}

Due to the shallow depth and large point spread function (PSF) of Herschel imaging (PSF FWHM $\sim$ 7-25''), confusion noise dominates the images. The cosmic FIR background, generated by faint unresolved sources, sets a limit below which individual sources cannot be detected. As a result, bright nearby objects blend together while faint ones are buried in the confusion noise, leaving FIR photometry of distant galaxies limited and noise-dominated \citep{PopeChary2008}.

Because of this severe blending, it is challenging to identify and stack clean point sources directly from the science field. As a result, a high signal-to-noise empirical PSF cannot be constructed \citep[as is possible for Spitzer data in, e.g.,][]{Whitaker2014}. Instead, we adopt the calibration PSF of the asteroid Vesta\footnote{\url{https://www.cosmos.esa.int/web/herschel/ancillary-data-products}}, which provides an uncontaminated and high S/N model of Herschel's PSF shape. This ensures that the PSF model used in fitting and subtraction is accurate, thus improving deblending and flux recovery despite the large beam.
\replaced{For the field/bands where the Herschel data are heterogeneous and combine multiple scan directions and depths (i.e., GOODS-S PACS) the effective PSF varies across the mosaic. In these cases, we generate position-dependent PSFs that account for the position angles and the integration time of the telescope visit over the fields.}{When Herschel data are heterogeneous and combine multiple scan directions and depths (i.e., GOODS-S PACS), the effective PSF varies across the mosaic. We construct a position-dependent PSF for each source location by rotating the Vesta calibration PSF to the orientation of each contributing scan, then co-adding the rotated copies weighted by the per-pixel integration time from each scan direction at that location. The result is a unique effective PSF that captures the local scan-pattern geometry, rather than a single global PSF averaged across the mosaic. This treatment is most important in regions where the GOODS-Herschel and PEP scan directions contribute unequally, since the effective PSF in those locations differs noticeably in orientation and shape from the global average.}
This improves the clean subtraction of nearby sources, thereby achieving more accurate flux estimates for faint sources that fall in the PSF wings of brighter sources. 
These Herschel PSFs are then fed into {\fontfamily{qcr}\selectfont MOPHONGO} to construct transfer kernels between the HST detection and Herschel images.

To further mitigate source blending, we perform forced photometry on the mosaics in the four CANDELS fields. Contamination from neighboring objects is modeled using HST NIR imaging as positional priors and FIR flux predictions from {\fontfamily{qcr}\selectfont Prospector} FUV-MIR SED modeling \citep{Leja2017} as flux priors. Incorporating these priors identifies more FIR-bright galaxies \citep{Lang2016} and thus removes their contamination with higher fidelity, thereby boosting signal-to-noise and improving flux measurements for faint sources that lie below the confusion limit.

In the Herschel photometric extraction, we use {\fontfamily{qcr}\selectfont T-PHOT} \citep{Merlin2015, Merlin2016}, incorporating FIR {\fontfamily{qcr}\selectfont Prospector} predictions and their uncertainties as Gaussian priors. 
Following the fitting approach described in Section 2.4 of \citet{Merlin2016}, these priors are incorporated into the {\fontfamily{qcr}\selectfont T-PHOT} fitting as a regularization term on the source fluxes, while the Herschel fluxes themselves remain free parameters. The prior information helps stabilize the solution in confusion-limited regimes by suppressing degenerate flux assignments among deblended sources when the Herschel data alone are insufficient, while allowing the data to dominate the fit for well-detected objects.

FIR predictions from {\fontfamily{qcr}\selectfont Prospector} FUV-MIR modeling are available for galaxies at $0.5<z<2.5$ above the 3D-HST mass completeness limit of $\log(M_*/M_{\odot}) = 9.3$ \citep{Tal2014}, which comprises $>$93\% of FIR detected sources (at least two Herschel filters detected) and 33\% of sources in our entire sample \citep{Leja2019b}. 
For galaxies without individual {\fontfamily{qcr}\selectfont Prospector} predictions, we adopt priors based on the median predictions of sources at similar redshift ($z_{\rm best}$) and MIPS 24$\mu$m flux. 

For sources outside the $0.5<z<2.5$ window studied in \citet{Leja2019b}, we assign priors by extrapolating the median {\fontfamily{qcr}\selectfont Prospector} observed-frame 24$\mu$m-to-Herschel flux ratios within that range. This approach, which \citet{Leja2019b} adopt with {\fontfamily{qcr}\selectfont Prospector-$\alpha$} modeling, fixes the parameters in the IR SED from \citet{DraineLi2007} to exactly match the inferred conversion factors ($L_{\mathrm{8}}$/$L_{\mathrm{IR}}$) from early Herschel studies \citep[e.g.,][]{Elbaz2011} in an attempt to have no redshift evolution.
 
With the imposed position and flux priors described above, we model the Herschel imaging for our sample in the 3D-HST catalog with {\fontfamily{qcr}\selectfont T-PHOT}. We use Herschel images that are registered to the reference HST image and rebinned to the same pixel scale (0\farcs18/pixel and 0\farcs36/pixel for PACS and SPIRE images, respectively) as input. In {\fontfamily{qcr}\selectfont T-PHOT}, 
the HST segmentation map is convolved with a transfer kernel to construct models for the low-resolution data (i.e., Herschel) for each source, with the flux of each source being left free to vary. The best-fit Herschel fluxes are then obtained by fitting these models to the observed Herschel data. While {\fontfamily{qcr}\selectfont T-PHOT} provides best-fit flux estimates for all sources, in this work we use the resulting source models to subtract contaminating neighbors. 


\subsubsection{Aperture photometry}

Using the cleaned images produced by subtracting the best-fit {\fontfamily{qcr}\selectfont T-PHOT} models of neighboring sources, we measure final Herschel fluxes via aperture photometry.
In brief, we construct a tile of a size of $\sim$9 times the PSF full-width half maximum (FWHM) for PACS100 and $\sim$7 times the FWHM for SPIRE350 centered on each source. 
We then perform aperture photometry of the central source in a radius of $\sim$0.5 PSF FWHM (e.g., 3\arcsec\ in PACS70, 12\arcsec\ in SPIRE350) in the cleaned {\fontfamily{qcr}\selectfont T-PHOT} image. 
The measured flux is corrected for the finite size of the PSF, determined by comparing the curve-of-growth of the reference PSF of the asteroid Vesta. For the PSF size adopted in this study, this correction is $\lesssim$ 10\%\footnote{12,14,15,6,4\% for PACS70, PACS100, PACS160, SPIRE250, SPIRE350, respectively}.
An additional correction is then applied to account for flux loss of 11--13\% due to high-pass filtering in PACS bands following \citet{Popesso2012}\footnote{13,12,11\% for PACS70, PACS100, PACS160, respectively}.

\begin{table}
\centering
\caption{\label{tab:mcfit} Best-fit $r$--$\sigma_{f}$ relation}
\tabletypesize{\normalsize}
\begin{tabular*}{\columnwidth}{@{\extracolsep{\fill}} lcc}
\hline\hline
Band & $a$ & $b$ \\
\hline
SPIRE250 & $-0.56$ & 2.11 \\
SPIRE350 & $-0.64$ & 2.62 \\
\hline
\end{tabular*}
\tablecomments{ \scriptsize The best-fit for the correction factor ($r$) as a function of flux uncertainties ($\sigma_f$) obtained in our Monte Carlo simulations, modeled as $\log r$=$a \times \log \sigma_f + b$, which is applied to the original flux errors in SPIRE bands.}
\end{table}


\begin{deluxetable*}{lllll}
\tablecaption{\label{tab:filters} Image sources}
\tabletypesize{\scriptsize}
\tablewidth{0pt}
\tablehead{
Field & Filters & Telescope/Instrument & Survey & Reference 
}
\startdata
COSMOS & $u$, $g$, $r$, $i$, $z$ & CFHT/MegaCam & CFHTLS &  \citet{Erben2009, Hildebrandt2009}\\
 & $B$, $V$, $r'$, $i'$, $z'$, 12 medium-band optical  & Subaru/Suprime-Cam & &  \citet{Taniguchi2007}\\
 &F606W, F814W   & HST/ACS & CANDELS & \citet{Grogin2011, Koekemoer2011}\\
 &$J1$, $J2$, $J3$, $H1$, $H2$, $K$ & KPNO 4m/NEWFIRM & NMBS & \citet{Whitaker2011} \\
 &$J$, $H$, $K_s$  &  CFHT/WIRCam & WIRDS & \citet{Bielbly2012}\\
 &$Y$, $J$, $H$, $K_s$ & VISTA & UltraVISTA & \citet{McCracken2012} \\
 &F140W & HST/WFC3 & 3D-HST & \citet{Brammer2012a}\\
 &F125W, F160W & HST/WFC3 &  CANDELS & \citet{Grogin2011, Koekemoer2011}\\
 &3.6, 4.5$\mu m$ & Spitzer/IRAC & SEDS & \citet{Ashby2013} \\
 &5.8, 8$\mu m$ & Spitzer/IRAC & S-COSMOS & \citet{Sanders2007}\\
 & 24$\mu m$ & Spitzer/MIPS & S-COSMOS & \citet{Sanders2007} \\
 & 100, 160$\mu m$ & Herschel/PACS & CANDELS-H, PEP & H. Inami (Priv. Comm) \\
 & 250, 350$\mu m$ & Herschel/SPIRE & CANDELS-H & H. Inami (Priv. Comm) \\
\noalign{\smallskip}
\hline
\noalign{\smallskip}
GOODS-N & $U$  & KPNO 4m/Mosaic & Hawaii HDFN & \citet{Capak2004}\\
 & $G$, $R_s$  &  Keck/LRIS & & \citet{Steidel2003}\\
 & F435W, F606W, F775W, F850LP   & HST/ACS & GOODS  &  \citet{Giavalisco2004} \\
 & $B$, $V$, $R_c$, $I_c$, $z'$ & Subaru/Suprime-Cam & Hawaii HDFN  &  \citet{Capak2004} \\
& F140W   & HST/WFC3 & 3D-HST &  \citet{Brammer2012a}\\
 & F125W, F160W  & HST/WFC3 & CANDELS &\citet{Grogin2011, Koekemoer2011}\\
 & $J$, $H$, $K_s$  & Subaru/MOIRCS & MODS &  \citet{Kajisawa2011}\\
  &3.6, 4.5$\mu m$ & Spitzer/IRAC & SEDS & \citet{Ashby2013} \\
&5.8, 8$\mu m$ & Spitzer/IRAC & GOODS &  \citet{dickinson2003}\\
 & 24$\mu m$ & Spitzer/MIPS & GOODS & \citet{dickinson2003} \\
 & 100, 160$\mu m$ & Herschel/PACS & GOODS-H, PEP & \citet{Magnelli2013}, H. Inami (Priv. Comm) \\
 & 250, 350$\mu m$ & Herschel/SPIRE & GOODS-H & \citet{Elbaz2011}, H. Inami (Priv. Comm) \\
\noalign{\smallskip}
\hline
\noalign{\smallskip}
GOODS-S &$U$, $R$  & VLT/VIMOS & ESO/GOODS &  \citet{Nonino2009}\\
& $U38$, $B$, $V$, $R_c$, $I$ & WFI 2.2m & GaBoDs & \citet{Hildebrandt2006, Erben2005}\\
& 14 medium bands & Subaru/Suprime-Cam & MUSYC & \citet{Cardamone2010}\\
 &F435W, F606W, F775W, F850LP   & HST/ACS & GOODS &  \citet{Giavalisco2004} \\
 &F606W, F814W, F850LP  & HST/ACS & CANDELS & \citet{Grogin2011, Koekemoer2011}\\
&F140W  & HST/WFC3 & 3D-HST &  \citet{Brammer2012a} \\
& F125W, F160W  & HST/WFC3 & CANDELS & \citet{Grogin2011, Koekemoer2011}\\
 &$J$, $H$, $K_s$  & VLT/ISAAC & ESO/GOODS, FIREWORKS &  \citet{Retzlaff2010}, \citet{Wuyts2008} \\
 & $J$, $K_s$ & CFHT/WIRcam & TENIS & \citet{Hsieh2012}\\
 &3.6, 4.5$\mu m$ & Spitzer/IRAC & SEDS & \citet{Ashby2013} \\
 &5.8, 8$\mu m$ & Spitzer/IRAC & GOODS &  \citet{dickinson2003} \\
  & 24$\mu m$ & Spitzer/MIPS & GOODS & \citet{dickinson2003} \\
  & 70, 100, 160$\mu m$ & Herschel/PACS & GOODS-H, PEP & \citet{Magnelli2013}, H. Inami (Priv. Comm) \\
 & 250, 350$\mu m$ & Herschel/SPIRE & HerMES & \citet{Oliver2012} \\
\noalign{\smallskip}
\hline
\noalign{\smallskip}
UDS &  $U$   & CFHT/MegaCam & & Almaini/Foucaud in prep.\\
 & $B$, $V$, $R_c$, $i'$, $z'$   & Subaru/Suprime-Cam & SXDS &  \citet{furusawa08} \\
 &F606W, F814W  & HST/ACS &  CANDELS & \citet{Grogin2011, Koekemoer2011}\\
 &F140W  & HST/WFC3 & 3D-HST &  \citet{Brammer2012a} \\
 &F125W, F160W  & HST/WFC3 & CANDELS & \citet{Grogin2011, Koekemoer2011} \\
 &$J$, $H$, $K_s$ &  UKIRT/WFCAM & UKIDSS DR8 & Almaini in prep. \\
  &3.6, 4.5$\mu m$ & Spitzer/IRAC & SEDS & \citet{Ashby2013} \\
&5.8, 8$\mu m$ & Spitzer/IRAC & SpUDS & Dunlop in prep.\\
 & 24$\mu m$ & Spitzer/MIPS & SpUDS & Dunlop in prep. \\
 & 100, 160$\mu m$ & Herschel/PACS & CANDELS-H & H. Inami (Priv. Comm) \\
 & 250, 350$\mu m$ & Herschel/SPIRE & CANDELS-H & H. Inami (Priv. Comm) \\
\enddata
\tablecomments{UV--8$\mu m$ list is taken from \citet{Skelton2014}}
\end{deluxetable*}


\subsubsection{Error budget}\label{sec:errorbudget}

One source of uncertainty that may not be properly accounted for in our adopted scheme of post-{\fontfamily{qcr}\selectfont T-PHOT} aperture photometry is the uncertainties of flux priors. The flux prior uncertainty will become increasingly important at the longest wavelengths where resolution is poorest and blending is most severe. 

Although the neighbor-subtracted residual image is derived from the single best-fit {\fontfamily{qcr}\selectfont T-PHOT} solution, this residual does not automatically propagate the uncertainty in the flux priors of neighboring sources into the cleaned image. To assess the impact of neighbor flux uncertainty on our aperture photometry---and to evaluate whether our measured flux uncertainties are realistic---we run Monte Carlo simulations in which we randomly perturb the fluxes of nearby sources following the Gaussian probability distribution of their flux priors as described in Section \ref{subsec:ForcedPhotometry}. We then build the corresponding residual map where the nearby sources are subtracted and the flux density of the central source is measured. We repeat the procedure 30 times for each of the $\gtrsim$ 100 sources for each field/band, which are selected to be uniformly distributed in SNR (0 $<$ SNR $<$ 20 in PACS100). 

The standard deviation of the output flux distribution obtained from the above procedure relative to the original flux errors indicates that our SPIRE uncertainties are underestimated, and thus corrected as follows.

We find a clear anti-correlation of this ratio ($r$) with the original SPIRE band flux uncertainties ($\sigma_f$), whereas no significant trend is found with flux or SNR.
The best-fit for the ratio $r$ as a function of flux uncertainty, modeled as $\log r$=$a \times \log \sigma_f + b$, is adopted as the scale factor needed to be applied to the original flux errors in SPIRE. Table \ref{tab:mcfit} lists the best-fit parameters. The correction factor increases with wavelength, from $r \sim$ 3 in SPIRE250 to $r \sim$ 6 in SPIRE350 at $\sigma_f$=1 mJy, but no significant variations are seen between the four fields. 

We perform a second test of the updated flux uncertainties in all PACS and SPIRE bands. We first check the residual maps to confirm that the background peaks at zero, and then compare the width of the residual flux distribution to our inferred errors. Whereas the SPIRE filters are fine, the residual widths in the PACS filters are consistently a factor of two larger than the reported noise.  This indicates that the PACS flux uncertainties remain underestimated. We thus scale the PACS error using the width of the residual map divided by the original flux uncertainties as the factor. Together, these corrections are comparable to the factor of $\sim5$ derived from simulation-based uncertainty calibrations in the COSMOS catalog by \citet{Jin2018}, with our analysis providing a complementary empirical basis for such an adjustment. 

While we do not preserve the full covariance between closely separated sources, our reduction procedure naturally elevates uncertainties for sources near bright neighbors. After {\fontfamily{qcr}\selectfont T-PHOT} subtracts adjacent sources from the FIR map, the local background and uncertainty are re-estimated from the residual image, meaning imperfect neighbor subtraction is directly propagated into the final per-source flux uncertainty. To quantify this, we compare the fractional flux uncertainties ($\sigma_F / F$) of intermediate-flux sources in GOODS-South with and without a bright neighbor within one FWHM, binning by flux to control for any native brightness difference between the two populations. We find fractional uncertainty boost factors of $\sim$1.42$\times$ across PACS bands (1.50$\times$ at 100$\mu$m, 1.34$\times$ at 160$\mu$m) and $\sim$1.11$\times$ across SPIRE bands (1.15$\times$ at 250$\mu$m, 1.07$\times$ at 350$\mu$m), confirming that sources in crowded environments carry appropriately larger uncertainties.

These corrections ensure conservative flux uncertainties that are inflated appropriately considering the low-resolution Herschel images. We caution that catastrophic failures in the flux priors may occur in individual cases. However, because the T-PHOT solution is ultimately constrained by the pixel-level noise properties of the low-resolution image, sources with intrinsic SNR$<$1 are driven toward fluxes consistent with the image noise rather than the prior model values. As such, the methodology is relatively insensitive to prior mismatches at flux levels below the detection threshold. The extracted Herschel FIR photometry is then combined with the 3D-HST photometric catalogs \citep{Skelton2014} and the 24$\mu$m Spitzer fluxes from \citet{Whitaker2014}, forming a long-wavelength (0.3 - 350\micron) master catalog, 3D-Herschel. Table \ref{tab:filters} lists the photometric bands, telescope/instruments, and surveys covering the four fields in 3D-Herschel, with references available for further details. The resulting UV-to-FIR 3D-Herschel photometric catalogs for all four fields are publicly available on Zenodo (McNulty et al. 2026, \dataset[doi:10.5281/zenodo.20706940]{https://doi.org/10.5281/zenodo.20706940}).

\begin{figure*}[ht!]
\plotone{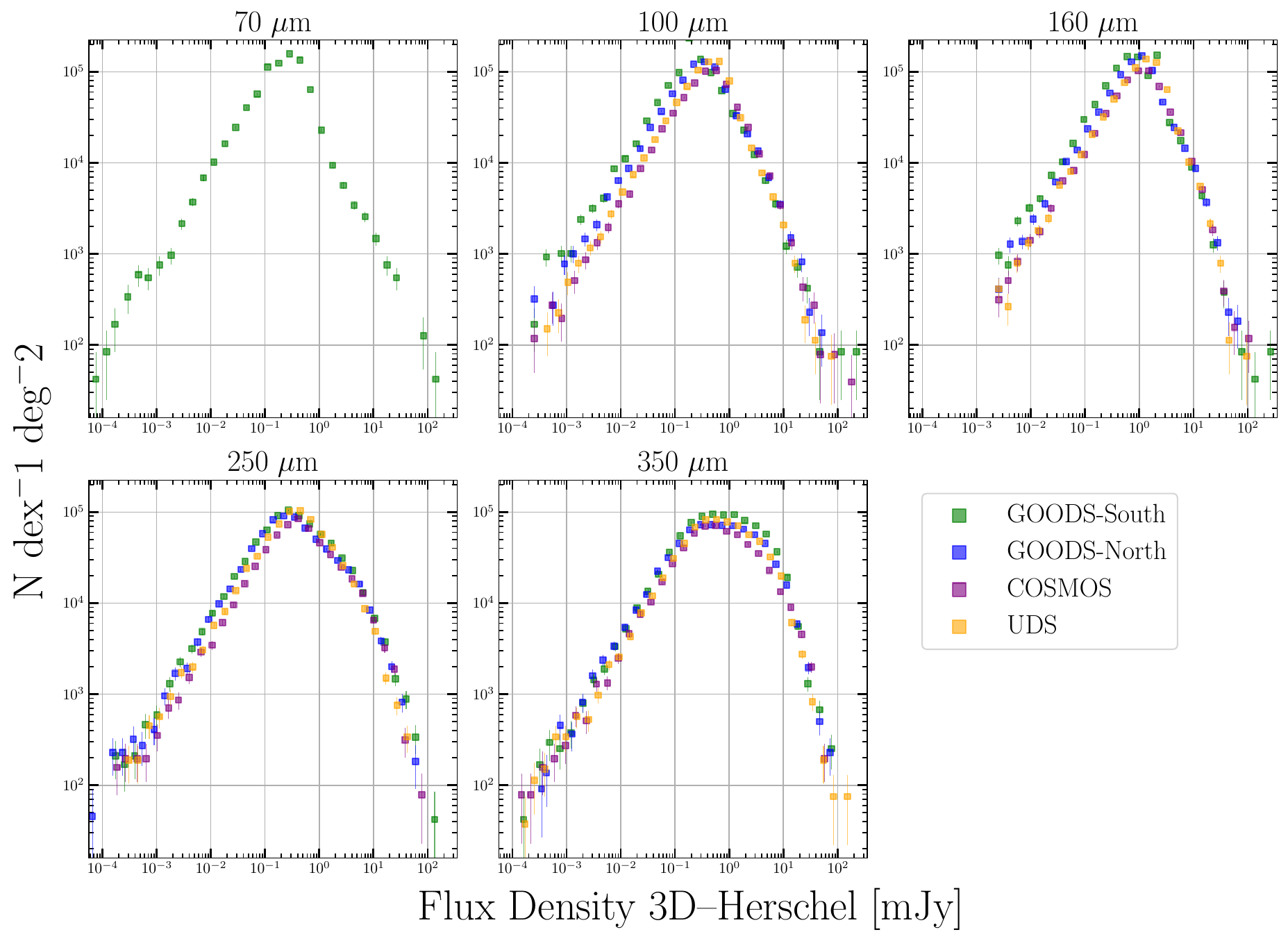}
\caption{Close relationship between the number counts of each 3D-Herschel field (binned by flux) demonstrates that the photometric estimations were consistent across all fields.
Error bars are determined as the fraction error of each bin ($ 1/ \sqrt{N} \times N \ dex^{-1} deg^{-2} $). 
\label{fig:galaxy counts}}
\end{figure*}

\section{Validation}  \label{sec:validation}

To verify that the 3D-Herschel photometry was correctly extracted, we 
cross-examine sources in each field against existing independent Herschel 
catalogs (Table~\ref{table:3D-Herschel}). Galaxy counts of the 3D-Herschel 
photometric catalogs are presented in Figure~\ref{fig:galaxy counts}, 
demonstrating that photometric estimates are consistent across all fields, 
with minor variations at the peak owing to field-to-field differences in 
FIR imaging depth. A full flux comparison and statistical validation are 
presented in Appendix \ref{app:validation}. With validated 3D-Herschel 
photometric catalogs, SED modeling with {\fontfamily{qcr}\selectfont 
Prospector} using the long-wavelength combined catalogs can be completed 
as described next.



\section{SED Modeling}  \label{sec:modeling}


We use the most recent and flexible version, {\fontfamily{qcr}\selectfont Prospector-$\beta$}, as the baseline of our SED modeling \citep{Wang2023}. This 17-parameter framework incorporates a nonparametric SFH, two-component dust attenuation, and variable dust emission.\deleted{enabling a more complete treatment of galaxy physics than earlier 14-parameter implementations such as {\fontfamily{qcr}\selectfont Prospector-$\alpha$} used by \citet{Leja2019b}.} To make the analysis computationally feasible for our large sample, we rely on an emulator -- a neural network trained to reproduce full {\fontfamily{qcr}\selectfont Prospector} outputs -- rather than recalculating spectra at each MCMC step. The extended list of parameters and their priors included in the emulator are listed in Table 1 of \citet{Mathews2023}. The emulator yields nearly identical posterior distributions for most parameters, but at a fraction of the computational cost, allowing the four fields to be modeled within weeks, as each object requires on average 10 minutes of compute time.

For continuity with past work, Section~\ref{subsec:3DHST model} first benchmarks the emulator against the simpler {\fontfamily{qcr}\selectfont Prospector-$\alpha$} framework applied to the 3D-HST catalog, before turning to our new 17-parameter modeling of the 3D-Herschel dataset in Section~\ref{subsec:Emulation}.




{\normalsize 
\begin{deluxetable*}{lcccc}
\tablecaption{Corresponding redshifts in their respective ranges with stellar mass limits \label{table:redshifts}}
\tablewidth{0pt} 
\tablehead{
\colhead{\normalsize Redshift Range} & 
\colhead{\normalsize Completeness Limit [log(M$_{\star}$/M$_{\odot}$)]} & 
\colhead{\normalsize \# Spectroscopic} & 
\colhead{\normalsize \# Grism} & 
\colhead{\normalsize \# Photometric}
}
\startdata
\addlinespace[2pt]
\normalsize 0.5 $<$ \textit{z} $<$ 0.8  & \normalsize 8 & \normalsize 955 & \normalsize 2,297 & \normalsize 5,778 \\
\addlinespace[2pt]
\normalsize 0.8 $<$ \textit{z} $<$ 1.2  & \normalsize 8.5 & \normalsize 822 & \normalsize 2,930 & \normalsize 6,660 \\
\addlinespace[2pt]
\normalsize 1.2 $<$ \textit{z} $<$ 1.8  & \normalsize 9 & \normalsize 359 & \normalsize 3,193 & \normalsize 6,314 \\
\addlinespace[2pt]
\normalsize 1.8 $<$ \textit{z} $<$ 2.5  & \normalsize 9.4 & \normalsize 135 & \normalsize 1,600 & \normalsize 3,400 \\
\addlinespace[2pt]
\hline
\addlinespace[2pt]
\normalsize Total: & & \normalsize 6.6\% & \normalsize 29.1\% & \normalsize 64.3\% \\
\addlinespace[2pt]
\enddata
\end{deluxetable*}
}


\subsection{3D-HST modeling fits} \label{subsec:3DHST model}

The full {\fontfamily{qcr}\selectfont Prospector} SED modeling code has been employed to fit the 3D-HST photometric catalogs (0.3 - 24\micron), which serve as our baseline comparisons. When modeling the IR emission, {\fontfamily{qcr}\selectfont Prospector} assumes energy balance by having dust attenuated in the UV-NIR re-emitted as IR radiation \citep{daCunha2008}, and the shape of the IR SED is defined by dust emission templates provided by \citet{DraineLi2007}. To address the lack of observational constraints in the FIR, the IR SED (rest-frame $\sim$4-1000\micron) is fixed such that the conversion from Spitzer/MIPS 24\micron\ to total IR Luminosity ($L_{\mathrm{IR}}$, calculated from 8-1000\micron) approximates the log-average of the \citet{DaleHelou2002} templates, as described by \citet{Wuyts2008}. For the UV-MIR fits, the three dust emission parameters defined by the \citet{DraineLi2007} models, $q_{\mathrm{PAH}}$, $U_{\mathrm{min}}$, and $\gamma$ are fixed to 2, 1, and 0.01, respectively.

\begin{figure}[ht!]
\centering
\includegraphics[width=\columnwidth,height=0.7\textheight,keepaspectratio]{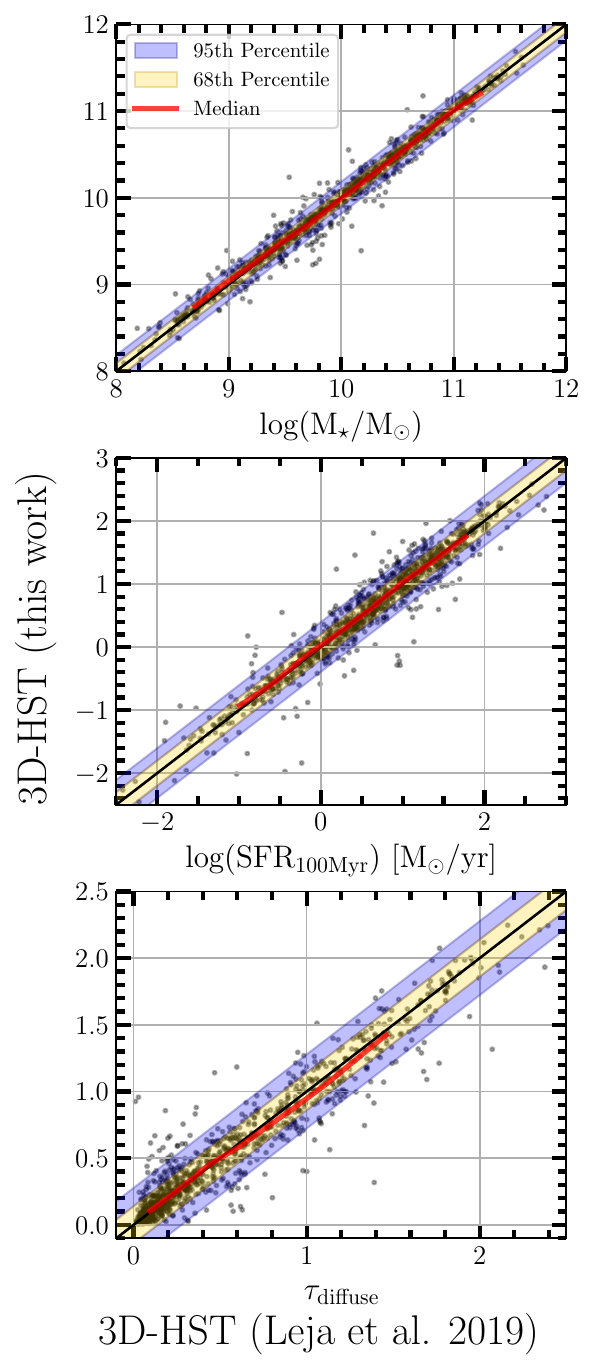}
\caption{Comparison of posterior medians inferred from full {\fontfamily{qcr}\selectfont Prospector} fits of the adjusted 3D-HST catalog (\citealt{Leja2019b}; x-axis) versus our {\fontfamily{qcr}\selectfont Prospector} emulation on the publicly released 3D-HST catalog (y-axis). Panels show (a) stellar mass, (b) SFR, and (c) diffuse‐dust attenuation.  The one-to-one line is in black and red lines show median offsets. The emulator reproduces direct‐fit estimates within their statistical uncertainties.
}
\label{fig:EmulatorVSfsps}
\end{figure}

\replaced{The earlier fits presented in \citet{Leja2019b} involve subtle variations in the treatment of the 3D-HST photometric catalog as compared to this work. 
During the assembly of the 3D-HST catalogs, as detailed in \citet{Skelton2014}, corrections to the zero-points were derived for each instrument and filter to 
optimize the absolute flux calibration. 
In lieu of applying these corrections, \citet{Leja2019b} instead inflated the flux errors by the equivalent amount for each photometric band, except for the space-based bands (HST and Spitzer), which inherently possess more stable zero-points. Instead, we leave the zero-point corrections as originally applied for the SED modeling and thus adopt the 3D-HST photometric catalogs wholesale. We refer the reader to \citet{Mathews2023} for a detailed comparison of the emulator results for the exact same 3D-HST photometric input.}{Unlike \citet{Leja2019b}, who inflated flux errors in lieu of zero-point corrections, we adopt the 3D-HST catalogs wholesale; see \citet{Mathews2023} for a detailed comparison of the emulator results for the exact same 3D-HST photometric input.}
In order to test for the impact of this change in approach relative to \citet{Leja2019b}, while also staying economical, we re-run the emulator on a subset of sources. 
Figure \ref{fig:EmulatorVSfsps} shows comparisons for six posterior medians for 1,000 randomly selected galaxies evenly distributed between the four fields, as presented in \citet{Leja2019b} (x-axis), to the emulation fits of the publicly released 3D-HST catalog (y-axis). We find no statistical difference in the inferred physical parameters within their quoted uncertainties; mean offsets are less than 0.1 dex for all presented parameters in Figure \ref{fig:EmulatorVSfsps}.

\subsection{3D-Herschel emulation fits} \label{subsec:Emulation}

To accommodate FIR SED modeling, we adopt slight modifications to the {\fontfamily{qcr}\selectfont Prospector} framework following \citet{Mathews2023}. In particular, the emulator now allows the three thermal dust emission parameters of \citet{DraineLi2007} -- previously fixed in fits without FIR constraints (Section~\ref{subsec:3DHST model}) -- to vary. These free parameters are the PAH mass fraction ($q_{\mathrm{PAH}}$), the minimum diffuse ISM radiation field strength heating the dust ($U_{\mathrm{min}}$), where $U$ is a dimensionless scaling factor, and the fraction of starlight ($\gamma$) exposed to radiation field strengths $U$ between $U_{\mathrm{min}}$ and $U_{\mathrm{max}}$, which we fix to $U_{\mathrm{max}} = 10^6$. 

Together with the 14 parameters of the baseline model and a fixed, galaxy-dependent redshift input, our physical model spans 17 parameters in total. Details of the prior distributions and emulator validation are provided in \citet{Mathews2023}, who also verify that the emulator shows no apparent bias and recovers posterior medians consistent with traditional SPS codes. This gives us confidence in using the emulator for this study, especially since modeling $\sim$41,000 galaxies demands the $10^{3}$–$10^{4}$ speed-up in runtime it provides.

\begin{figure*}[ht!]
\vspace{-1em} 
\hspace{1em} 
\includegraphics[width=\textwidth]{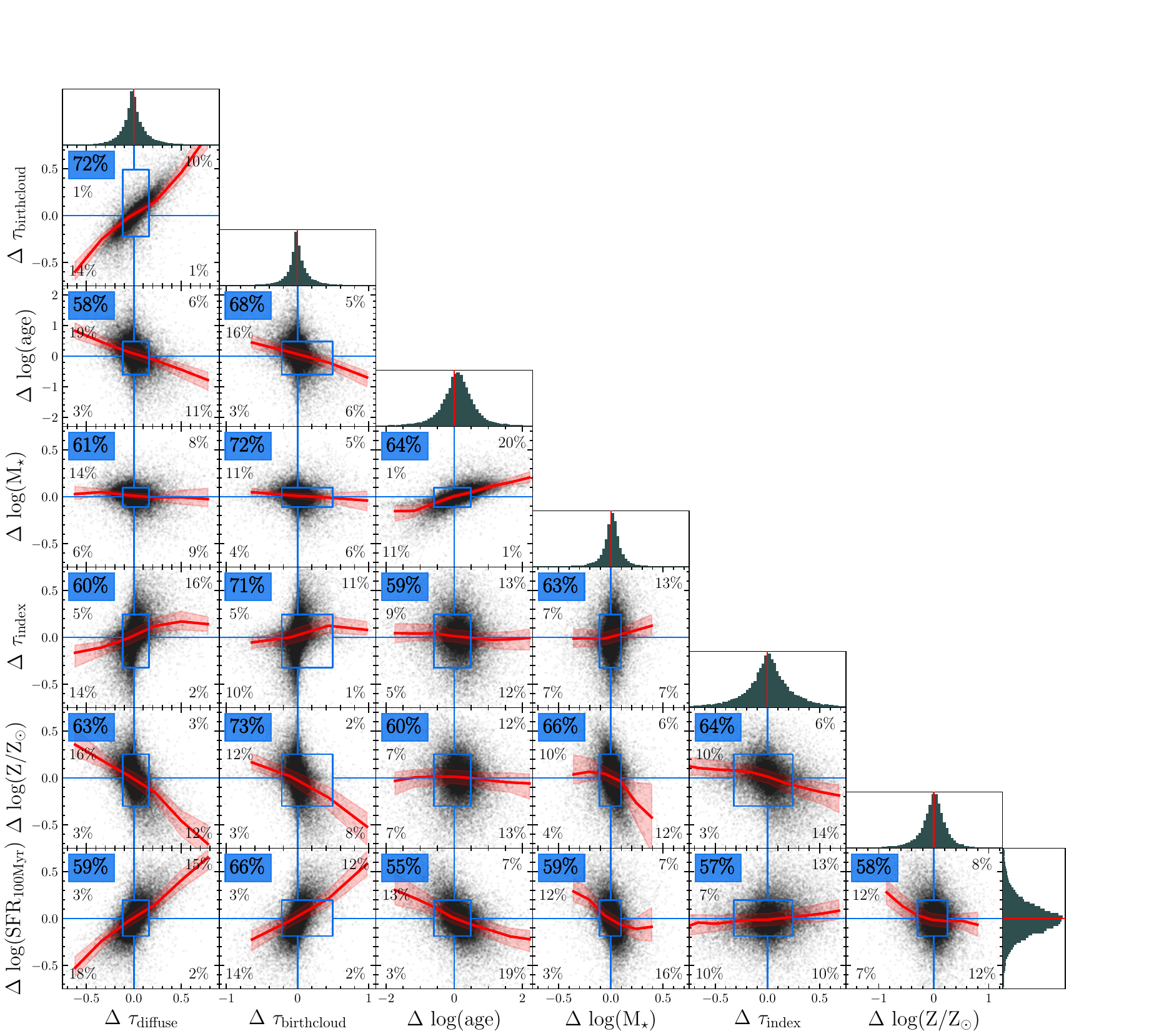}
\caption{Comparisons of fits with and without Herschel photometry, showing offsets (with - without Herschel) in 7 out of 17 of the inferred stellar parameters plotted against each other. Histograms of the offsets are shown above each column and at the end of the last row. Parameters (left to right, bottom row): diffuse dust attenuation ($\tau_{\mathrm{diffuse}}$), birth cloud dust attenuation ($\tau_{\mathrm{birthcloud}}$), mass-weighted age, stellar mass, diffuse dust index ($\tau_{\mathrm{index}}$), stellar metallicity, and SFR over the last 100 Myr bin (y-axis of bottom row). Blue boxes indicate the percent within 1$\sigma$, based on UV-MIR modeling; other indicated percentages refer to the population in the respective quadrant, excluding the 1$\sigma$ box.
\label{fig:cornerplot}}
\end{figure*}

\subsection{Sample selection}
\label{subsec:sample}

Within the four CANDELS fields we focus on herein, there are 166,767 sources in the catalog, whereas we focus hereafter on comparing results from \citet{Leja2019b, Leja2022} and thus model a subset of 41,387 galaxies. We adopt the 90\% stellar mass completeness limits following \citet{Tal2014}, as listed in Table~\ref{table:redshifts} along with the number of sources for each redshift range and the corresponding redshift type, leaving us with a mass-complete sample of 34,443 galaxies.

\section{Stellar Population Synthesis Modeling} \label{sec:results}

Using the UV-FIR photometry from the 3D-Herschel photometric catalogs, {\fontfamily{qcr}\selectfont Prospector} models the SED and infers the stellar parameters of each galaxy within a Bayesian framework.
\deleted{For each of the 17 free parameters, we adopt the median of the posterior distribution as the best estimate, with uncertainties given by the 16th and 84th percentiles.}

From the 17 physical parameters inferred from {\fontfamily{qcr}\selectfont Prospector}, we focus on a subset most likely to be affected by adding FIR constraints: star formation rate (SFR), surviving stellar mass ($M_{\star}$), diffuse and birth-cloud optical depths ($\tau_{\mathrm{diffuse}}$, $\tau_{\mathrm{birthcloud}}$), the slope of the attenuation law ($\tau_{\mathrm{index}}$), stellar age, and stellar metallicity. Notably, {\fontfamily{qcr}\selectfont Prospector} reports the surviving stellar mass, accounting for stellar mass loss from evolutionary processes (e.g., asymptotic giant branch (AGB) winds, supernovae), rather than the total mass formed. Throughout, 'stellar mass' refers to this surviving mass. We examine the offsets in these inferred SPS parameters between fits that include Herschel photometry versus those using photometry only up to 24\micron\ in Figure \ref{fig:cornerplot}.


\begin{figure*}[ht!]
\centering
\includegraphics[width=1.0\textwidth]{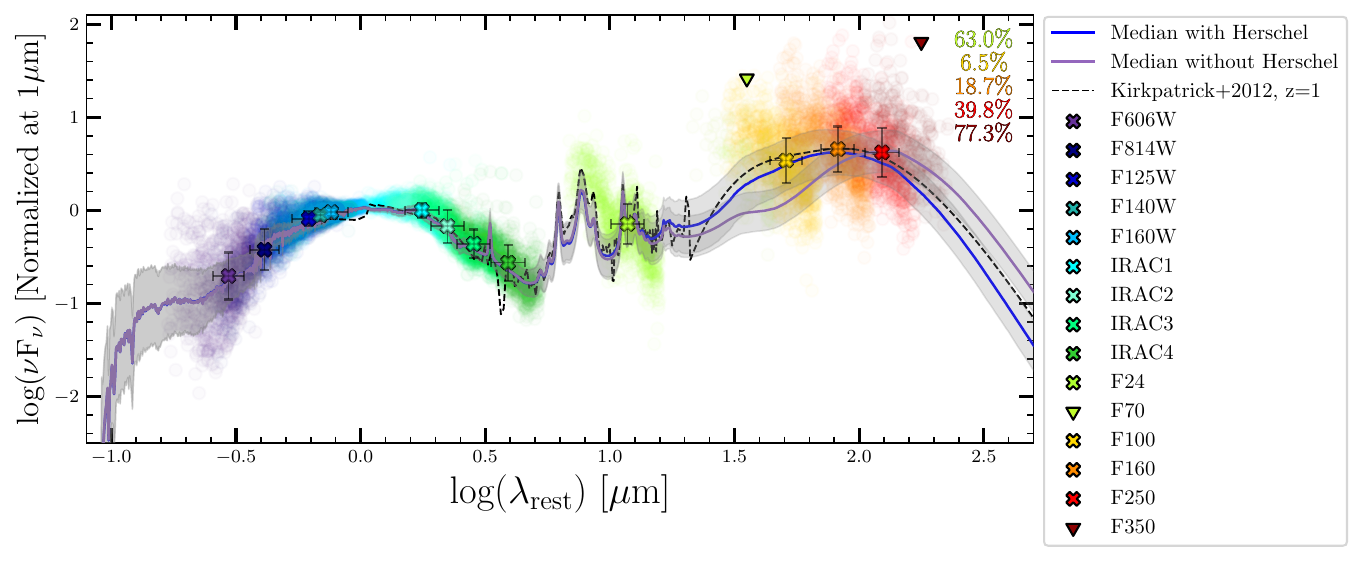}
\caption{Composite SEDs of the IR-bright sample from {\fontfamily{qcr}\selectfont Prospector} fits with Herschel FIR photometry added to the UV--MIR 3D-HST photometry (blue), compared to fits without Herschel (purple) and the \textit{z}$\sim$1 star-forming template of \citet{Kirkpatrick2012} (dashed). Each SED is normalized by the average flux at rest-frame 1\micron\ from the best-fit spectrum excluding Herschel photometry. The gray shaded region shows the $\pm1\sigma$ ensemble dispersion, computed as the median $\pm$ the median absolute deviation across all sources at each wavelength. Crosses and circles show median and individual photometry (SNR $>$ 3); inverted triangles are median upper limits for cases where $>$50\% of Herschel detections have SNR $<$ 3. The percentage of upper limits in each Herschel band is listed in the upper right. Redshifts range from 0.5-2.5, with 294 of the 1,118 sources having spectroscopic redshifts.
\label{fig:1composite}}
\end{figure*}


Overall, the parameters do not exhibit systematic offsets under the new modeling constraints; most histograms are centered around $\Delta = 0$ (red line). The inferred ages are an exception, showing a tendency toward older ages, though a majority ($\sim73\%$) of the sample remains within the 1$\sigma$ uncertainty. 

In each scatter subplot of Figure \ref{fig:cornerplot}, the 1$\sigma$ uncertainty box is outlined, corresponding to the median upper and lower 1$\sigma$ uncertainty of the respective parameter along that axis, taken from the fits with no Herschel photometry.\deleted{The individual 1$\sigma$ uncertainties are calculated by computing the distance between the median 84th and 50th percentile values for the upper uncertainty, and the median 50th and 16th percentile values for the lower uncertainty. The 84th, 50th, and 16th percentile values are taken from the fits with no Herschel photometry.} 
The blue box in the upper left corner of each subplot in Figure \ref{fig:cornerplot} shows the percentage of the population within the 1$\sigma$ region, indicating that most objects (55-69\%) deviate by less than 1$\sigma$ for the selected properties. However, the outliers show clear correlations when FIR constraints are added: dust attenuation ($\tau_{\mathrm{diffuse}}$ and $\tau_{\mathrm{birthcloud}}$) increases with higher SFRs and decreases stellar age. Based on these initial results, adding Herschel photometry does not introduce significant systematic offsets in the median inferences of key properties. Consequently, the inferred star-forming main sequence (SFMS) estimated from our sample of 34,443 objects is not systematically offset and remains in agreement with the previous fits presented in \citet{Leja2022}, showing an average deviation in the median (ridge) SFR of 0.1$\pm$0.07 dex and no greater than $\sim$0.25 dex for a fixed stellar mass. We defer a more in-depth discussion of the SFMS to Section \ref{subsec:SFMS}.

Herschel imaging is intrinsically low-resolution, and even after the deblending, a majority of measurements in our mass-limited sample are upper limits ($\gtrsim$ 90\%). Nevertheless, these upper limits still provide meaningful constraints in the modeling. For example, at low signal-to-noise (SNR $\approx 1$ for all 100, 160, 250, and 350\micron\ observations), the median uncertainties in log(SFR), $\mathrm{log(Z/Z_{\odot}})$, and $\tau_{\mathrm{diffuse}}$ are reduced by 0.06, 0.09, and 0.08 dex, respectively, relative to fits without Herschel photometry. 

However, to enable a more robust assessment, a high-quality sample of Herschel detections is required. In particular, such a sample is needed to quantify how reliably 24$\mu$m photometry traces the IR luminosity ($L_{\mathrm{IR}}$) and the stellar parameters dependent on $L_{\mathrm{IR}}$.


{\normalsize 
\begin{deluxetable*}{lrr}
\tablecaption{Percent of IR-bright sample in mass bins \label{table:IRbright}}
\tablewidth{0pt} 
\tablehead{
\colhead{\normalsize Mass Range [log($M_{\star}/M_{\odot}$)]} & 
\colhead{\normalsize Number of Galaxies} & 
\colhead{\normalsize Percent of Mass-Limited Population}
}
\startdata
\addlinespace[3pt]
\normalsize 9.0 $\leq$ log($M_{\star}/M_{\odot}$) $<$ 9.75  & \normalsize 17 & \normalsize 0.12\% \\
\addlinespace[3pt]
\normalsize 9.75 $\leq$ log($M_{\star}/M_{\odot}$) $<$ 10.5  & \normalsize 384 & \normalsize 4.43\% \\
\addlinespace[3pt]
\normalsize 10.5 $\leq$ log($M_{\star}/M_{\odot}$) $<$ 11.25 & \normalsize 653 & \normalsize 16.80\% \\
\addlinespace[3pt]
\normalsize 11.25 $\leq$ log($M_{\star}/M_{\odot}$) & \normalsize 64 & \normalsize 21.33\% \\
\addlinespace[3pt]
\enddata
\end{deluxetable*}
}

\begin{figure*}
\plotone{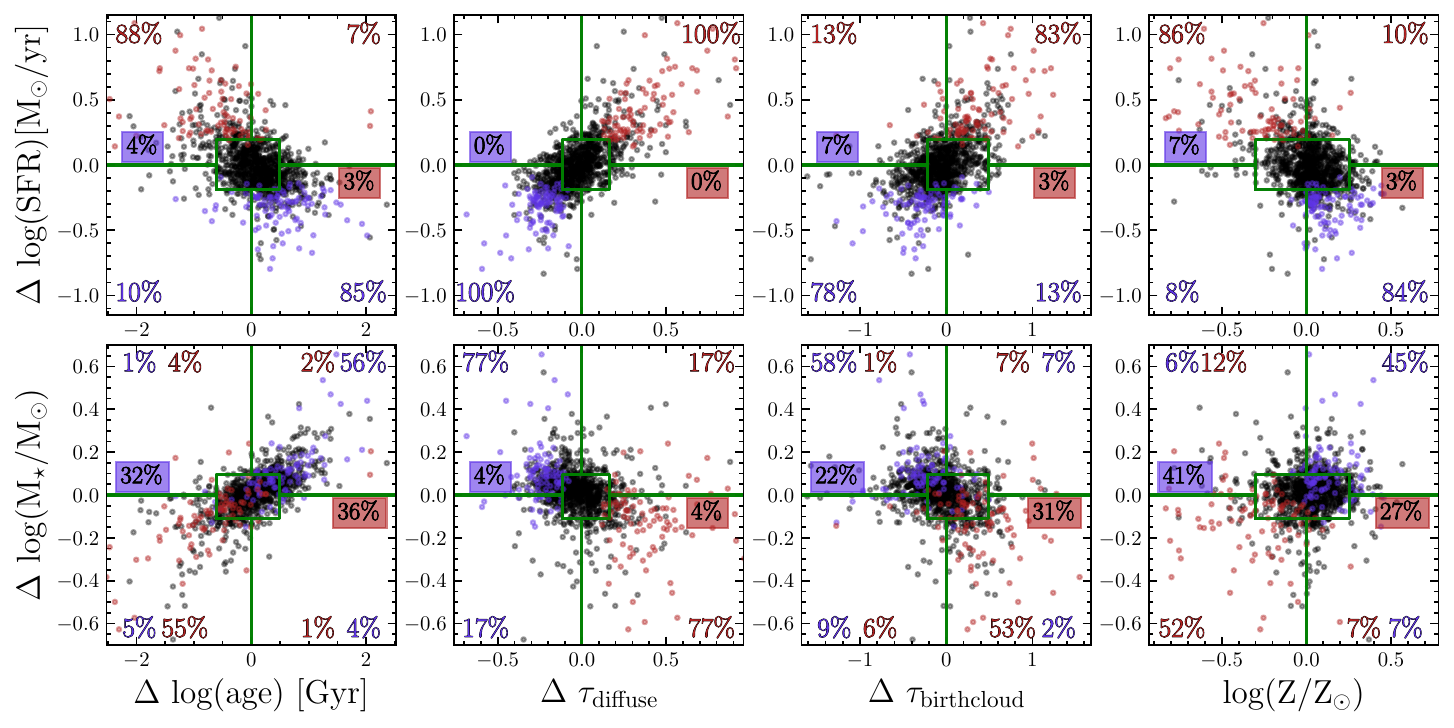}
\caption{Offsets (with - without Herschel) in star formation rate (top row) and stellar mass (bottom row) as a function of mass-weighted age, diffuse dust, birthcloud dust, and stellar metallicity (left to right). Objects are color-coded as dustier and more star-forming (red) or less dusty and less star-forming (purple). 
IR-bright galaxies modeled without FIR constraints are more poorly inferred than the full sample; the more dust-obscured, star-forming objects tend to be younger and more massive, while the less dusty population shows the opposite trend. Percentages of each subpopulation in each quadrant is reported in the corresponding color and corner, while the percentage within the 1$\sigma$ box is indicated in the colored text box.
\label{fig:ADM degen}}
\end{figure*}

\section{Changes in Inferred Parameters for Galaxies with Significant Herschel Detections} \label{sec:IR bright}


To construct an optimal sample of galaxies that are both well-imaged by Herschel and most likely to be affected by the new FIR constraints (i.e., IR-bright, typically meaning highly star-forming), we select sources with SNR $\geq$ 3 and a 'use' flag of 1 in at least two Herschel bands (PACS100, PACS160, SPIRE250, and SPIRE350) In addition, we require a MIPS 24$\mu$m detection (SNR $\geq$ 3), since the previous fits relied on the observed-frame 24$\mu$m flux as a proxy for the FIR.

Approximately 33\% of this population exhibit poor-fitting FIR SEDs, often due to bright interlopers contaminating the photometry of low-mass galaxies and producing overestimated fluxes in either PACS100 or PACS160. 
To remove these cases, we require $\chi^2 < 3$ in both PACS100 and PACS160 when comparing model fluxes to the deblended photometry. The final IR-bright sample consists of 1,118 objects. The fractional contribution of the IR-bright population to the total sample, as a function of stellar mass, is listed in Table \ref{table:IRbright}. Figure \ref{fig:1composite} shows the composite SEDs of this sample inferred from fits including Herschel photometry (blue) and from fits without Herschel (purple).

\deleted{To construct the composite, each SED is normalized by the average flux at rest-frame 1\micron\ from the best-fit spectrum excluding Herschel photometry, and the median spectrum is then calculated. The composites are supported by Herschel observations at 70, 100, 160, 250, and 350\micron\, as well as HST and Spitzer photometry from the 3D-HST photometric catalogs. For each object, the photometry is shifted to the rest frame, and median values are plotted as large X's, with standard deviations shown for filters in which fewer than 50\% of the data are upper limits. The fraction of upper limits in each Herschel filter is indicated in the top right corner, and for bands with more than 50\% upper limits, the limits are taken as three times the error and plotted as downward-pointing triangles.}

The composite SEDs are overplotted with the \textit{z}=1 star-forming template from \citet{Kirkpatrick2012}. The FIR spectrum of the 3D-Herschel IR-bright composite closely follows the shape of this template, in contrast to the fits without Herschel data, underscoring the robustness of our high-quality sample. We present this sample in greater detail, along with its placement on the SFMS, in the following section.

\subsection{Age-dust-metallicity degeneracy}

Dust attenuation is inferred from UV-optical-NIR starlight, and in the absence of FIR measurements, {\fontfamily{qcr}\selectfont Prospector} estimates the re-radiated dust emission by assuming energy balance. Extending photometric constraints into the FIR with Herschel are therefore critical for improving obscured SFR and related $\tau$ estimates, thus reducing degeneracies. These degeneracies arise from uncertainties in dust attenuation, which propagate into stellar age and metallicity estimates, since all three parameters redden the UV/optical SED. Figure \ref{fig:ADM degen} illustrates the effect by showing offsets in the mass-weighted age, optical depth of diffuse dust, and stellar metallicity as a function of the offsets in the SFR (top row) and stellar mass (bottom row). 

We separate the sample into two sub-populations: objects with higher diffuse dust attenuation and star formation (red) and those with lower values (purple) in their updated estimates, reflecting the critical role of Herschel data in reducing uncertainties in obscured SFR and $\tau_{\mathrm{diffuse}}$. To evaluate the significance of the offsets beyond the 1$\sigma$ box of our red and blue sub-populations, we include only objects whose updated uncertainties fall outside the range of their previous estimates. The fraction of each sub-population located in a given quadrant is reported in the corresponding corner, while the percentage within the 1$\sigma$ box is indicated in the respective colored box.

Relative to the mass-complete sample -- in which 24\micron\ serves as a robust proxy for FIR dust emission (with $>$60\% of objects contained within the 1$\sigma$ box; Figure \ref{fig:cornerplot}) -- the IR-bright sample shows modestly lower agreement: 73\% of stellar mass estimates and 66\% of SFR estimates remain consistent with the UV-MIR fits, compared to 78\% and 70\% for the overall population of 34,443 galaxies. The offsets shown in Figure \ref{fig:ADM degen} do not indicate substantial systematic biases in dust estimates based on MIR analogs: 90 objects are inferred to be significantly dustier and more star-forming, while 95 are less dusty and less star-forming, together comprising $\sim$8\% of the IR-bright sample.

The 1$\sigma$ boxes in Figure \ref{fig:ADM degen} provide a visual summary of the typical uncertainties in the baseline catalog and help visualize how the offsets correlate, but they do not robustly quantify the significance of the differences on an object-by-object basis. Here, we compute the uncertainty-normalized residuals for each source,
\begin{equation}
\frac{\Delta x}{\sigma_{\Delta x}} = \frac{x_{\mathrm{H}} - x_{\mathrm{noH}}}{\mathrm{\sqrt{\sigma^2_{H} + \sigma^{2}_{noH}}}}
\label{eq:flux_resid_param}
\end{equation}
where the subscripts H and noH denote values obtained with and without Herschel constraints. We find that stellar mass, SFR, and $\tau_{\mathrm{diffuse}}$ exhibit greater tension:$~$43\% of galaxies have $\vert\frac{\Delta x}{\sigma_{\Delta x}}\vert > 1$ and$~$15\% have $\vert\frac{\Delta x}{\sigma_{\Delta x}}\vert > 2$, exceeding Gaussian expectations. In contrast, age, $\tau_{\mathrm{birthcloud}}$, and metallicity, show a fewer percent of large residuals, more consistent with Gaussian-distributed uncertainties. \replaced{By directly constraining obscured SFR and dust emission, we obtain more reliable estimates of the degenerate parameters. Consequently, dustier, more star-forming systems are inferred to have younger ages, lower stellar masses, and reduced metallicities, while the opposite holds for less dusty and star-forming galaxies. Many of the offsets outside the 1$\sigma$ box are driven by $\tau_{\mathrm{diffuse}}$, likely reflecting the limitations of earlier $\tau$ estimates derived from models lacking FIR data. In the absence of FIR constraints, a fixed FIR color was used to match the log-averaged \citet{DaleHelou2002} templates, which either over- or underestimates the dust content in more than half of the IR-bright sample.}{In the absence of FIR constraints, the fixed log-averaged \citet{DaleHelou2002} template either over- or underestimates dust content for the majority of the IR-bright sample, with offsets in $\tau_{\mathrm{diffuse}}$ driving most of the scatter outside the 1$\sigma$ box.}



\begin{figure*}[ht!]
\centering
\includegraphics[width=1.0\textwidth]{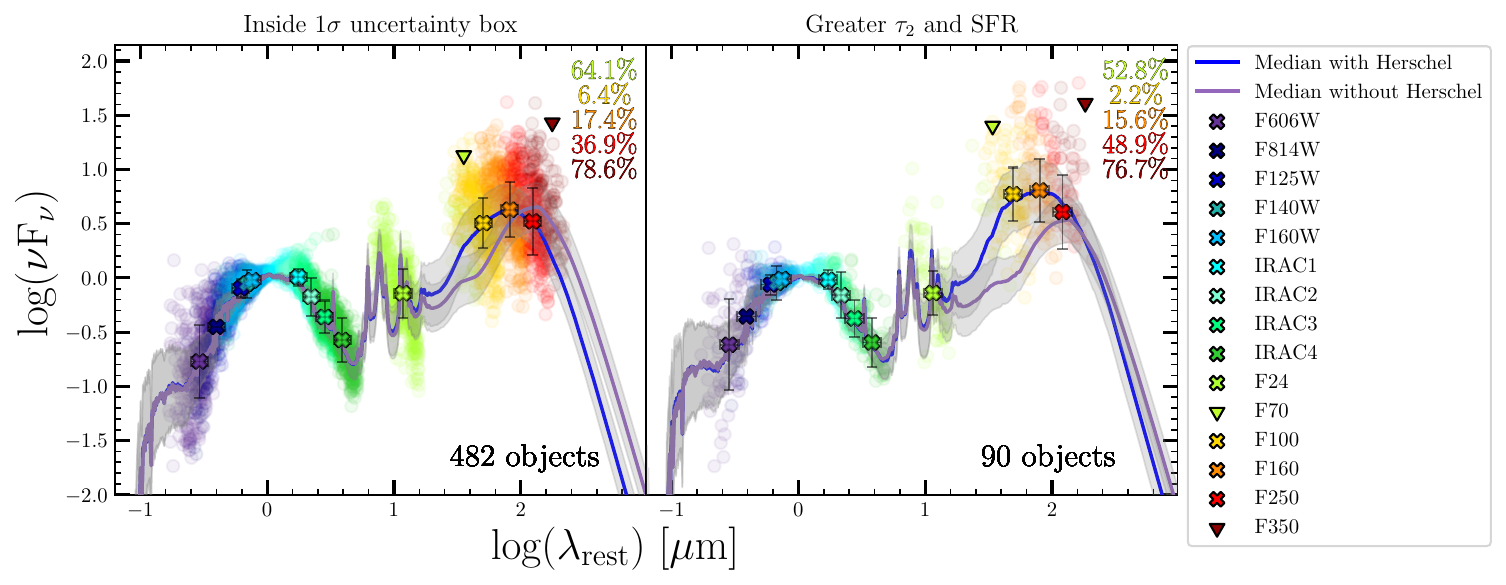}
\caption{Composite SEDs with (blue) and without Herschel photometry (purple), normalized at 1$\mu$m. The gray shaded region represents the $\pm1\sigma$ dispersion of the SED ensemble, computed as the median $\pm$ the median absolute deviation across all sources at each wavelength. The composites in the left panel are constructed from sources with diffuse dust and SFR offsets within their original 1$\sigma$ uncertainties, while the right panel is created from IR-bright objects that are significantly dustier and more star-forming under the new constraints. The percent of upper limits in each Herschel band is listed in the top right.
\label{fig:composites}}
\end{figure*}


To highlight the limitations of fixed SED templates used in fits without FIR constraints, Figure \ref{fig:composites} presents composite SEDs for the IR-bright sample: objects whose offsets remain within 1$\sigma$ (left panel) and those that are significantly dustier and more star-forming (right panel). These composites are constructed following the same procedure as in Figure \ref{fig:1composite}, as described in the caption of Figure \ref{fig:1composite}.

For objects where modeling up to the MIR was sufficient to estimate dust content and obscured star formation (left panel), the FIR shapes of the two fits nonetheless differ.
Incorporating Herschel constraints allows the FIR spectrum to vary freely, producing dust bumps that consistently peak at shorter wavelengths than in the previous fits. This trend also holds for the dustier, more star-forming systems shown in the right panel. The FIR dust bump is set by three dust emission parameters treated as free in our modeling, each directly linked to the emitting dust temperature, discussed further in Section \ref{DustParams}. Despite systematic differences in the FIR dust bump shape and consistently warmer peaks in the left panel, the total IR luminosity (3-1100$\mu$m) is not affected, consistent with the SFR and dust attenuation offsets of these objects.

In the right panel of Figure \ref{fig:composites}, the Herschel-informed composite SED exhibits substantially greater FIR emission than the fits that underestimated $\tau_{\mathrm{diffuse}}$ and SFR. For these 90 objects alone, the calculated $L_{\mathrm{IR}}$ is $0.3 \pm 0.25$ dex greater in models inferred with Herschel FIR photometry. Because FIR flux, and by extension IR luminosity, is a key proxy for the obscured SFR -- which dominates the star formation budget above log($M_{\star}$/$M_{\odot}$) $\sim$ 9.4 \citep{Whitaker2017} -- this difference is significant. Although a comparable number of objects show increases and decreases in SFR, the SFMS provides a more meaningful framework for evaluating how the IR-bright population is affecting well-studied scaling relations.


\begin{figure*}[ht!]
\includegraphics[width=1.0\textwidth]{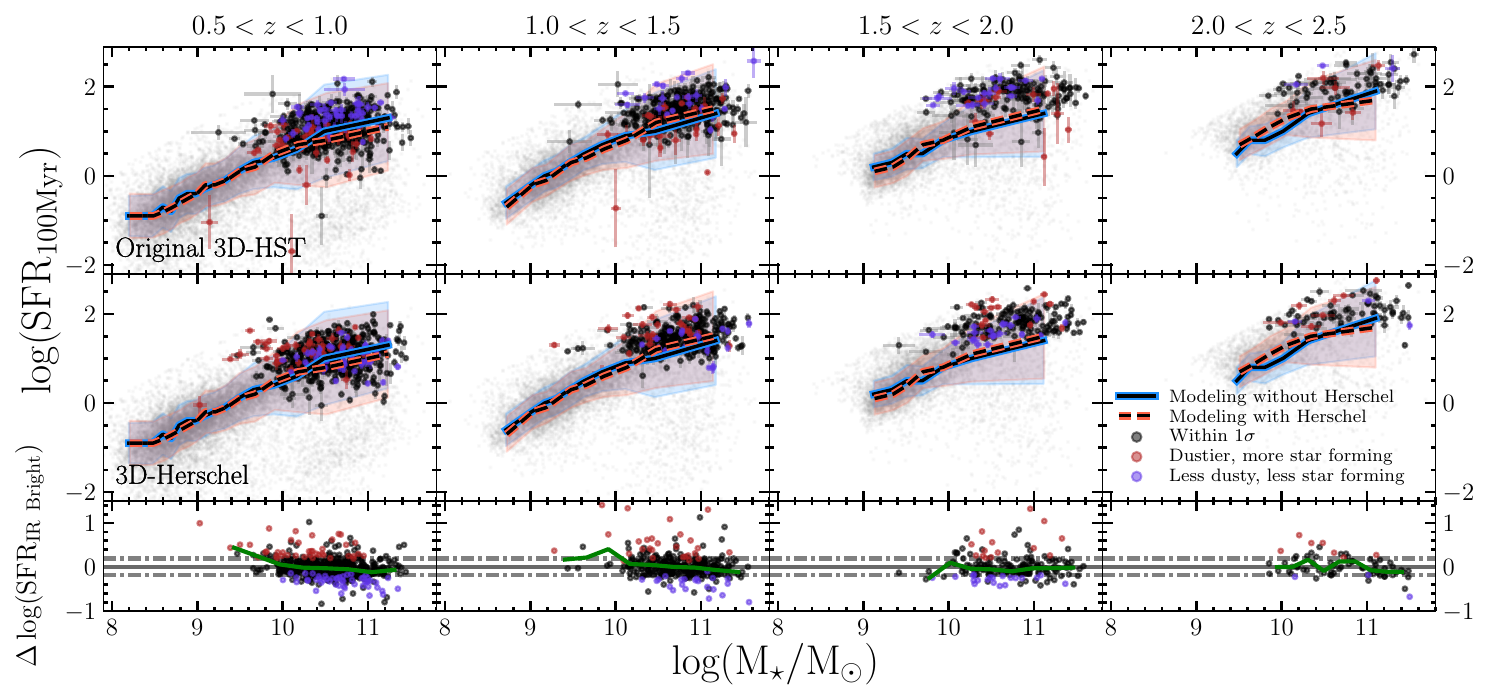}
\caption{
IR-bright population (circles) overplotted on the SFMS of the mass-limited sample, inferred from UV-MIR fits (blue line) and from Herschel-constrained fits (red dashed line). The top row shows SFRs and stellar masses from UV-MIR fits, whereas the middle row shows the same population from UV-FIR fits, and the bottom row shows SFR differences between the two sequences (Herschel -- UV-MIR). Objects are color-coded as in Figure \ref{fig:ADM degen}: significantly dustier and more star-forming (red) or less dusty and less star-forming (purple).
}
\label{fig:IRbright_SFsequence}
\end{figure*}


\subsection{Star-Forming Main Sequence}
\label{subsec:SFMS}

One of the most robustly measured relations in galaxy evolution is the SFMS: the tight correlation between stellar mass and SFR that traces where typical galaxies form the bulk of their mass, first identified by \citet{Noeske2007}. The existence of this relation indicates that most star-forming galaxies grow in a relatively steady fashion rather than through stochastic bursts, at least out to $z\sim3$. Given its central role in our understanding of galaxy growth, the SFMS has been characterized across a wide redshift range and with diverse methodologies \citep[e.g., see synthesis in][]{Speagle2014}, yielding variations in its slope and normalization \citep[e.g.,][]{Elbaz2007, Daddi2007} as well as scatter \citep{Whitaker2012,Speagle2014,Schreiber2018}. Later work revealed that the relation is not strictly linear: at high masses, the sequence flattens, introducing curvature \citep{Whitaker2014, Lee2015}. 

While subsequent studies have continued to refine measurements of the SFMS \citep[e.g.,][]{Abramson2015, Schreiber2015, Tomczak2016, Kurczynski2016, Popesso2019, Boogaard2018}, observations have generally shown a persistent $\sim$0.3 dex offset towards higher SFRs relative to predictions from cosmological simulations \citep{Furlong2015, Dave2019, Nelson2021}. \citet{Leja2022} recently reconciled this discrepancy using the same 3D-HST photometric catalogs employed in previous works \citep[e.g.,][]{Whitaker2014}, but with more sophisticated SPS modeling. Nonetheless, in the absence of direct FIR constraints, uncertainties in the underlying modeling assumptions remain. 

In Figure \ref{fig:IRbright_SFsequence}, we show the SFMS for our full sample of 34,443 galaxies, comparing {\fontfamily{qcr}\selectfont Prospector} fits that include Herschel photometry (red line) to those using UV-MIR data (blue line). The SFRs from {\fontfamily{qcr}\selectfont Prospector} are averaged over the most recent 100 Myr, rather than the instantaneous values, consistent with the timescale probed by UV+IR SFR indicators. We define the SFMS by its mode, or \textit{ridgeline}, following recent efforts to avoid the need for explicit quiescent/star-forming separation.\deleted{The mode more accurately identifies where most star formation occurs, appearing as the steep ridge on a contour map, while largely excluding quenched galaxies \citep{RenziniPeng2015}.} \replaced{\citet{Leja2022} estimated the ridgeline as the density peak in the SFR-mass plane using a flexible neural network (a normalizing flow). In this work, however, we adopt a similar but robust approach by directly estimating the mode by first discretizing the continuous values to the nearest 0.1, effectively computing the histogram mode with a bin width of 0.1.}{We estimate the ridgeline as the histogram mode with bin width 0.1 dex, similar to the density-peak approach of \citet{Leja2022}.}

The SFMS derived from Herschel-constrained fits and that recovered from UV–MIR–only ($\leq$24,$\mu$m) fits coincide closely and are not systematically offset across redshift bins, deviating by 0.1$\pm$0.07 dex on average (and no more than 0.25 dex in any stellar mass bin). This agreement demonstrates that rest-MIR photometry alone can robustly recover galaxy SFRs across a wide range of properties. In Figure \ref{fig:IRbright_SFsequence}, the IR-bright population is color-coded according to whether objects became significantly dustier and more star-forming (red), less dusty/star-forming (purple), or largely unchanged (black), as defined in Figure \ref{fig:ADM degen}. The bottom row shows the difference in the SFR between the two sequences (Herschel --  UV-MIR) as a function of the stellar mass inferred from the FIR-constrained modeling.

Adding FIR data does not substantially change the overall SFMS of IR-bright objects, although catastrophic under- or over-estimations of SFRs occasionally occur for individual cases, with offsets reaching up to an order of magnitude.
Galaxies that became dustier and more star-forming in the updated modeling (red) exhibit the largest discrepancies in comparison to the less dusty, less star-forming population (purple). Some objects with underestimated SFRs were previously inferred to lie in the green valley or near the quiescent population, while those with overestimated SFRs were often placed among the most actively star-forming galaxies in their mass bin. Between 0.5 $<$ \textit{z} $<$ 1.5, underestimates are more common at lower stellar masses, whereas overestimates are more frequently associated with higher-mass galaxies.

Nevertheless, MIR-limited fits recover the SFRs of the best Herschel-detected, FIR-bright galaxies with reasonable accuracy: $\sim$70\% of the IR-bright sample fall within 1$\sigma$ of the one-to-one relation, \added{while stellar masses are reliable for $\sim$80\% of galaxies.} The remaining outliers underscore the limitations of estimating the full IR luminosity from UV-MIR data with fixed templates alone, as discussed further in Section \ref{subsec: LIR}.

\subsection{$F_{\mathrm{TIR}}$ and $F_{\mathrm{7.7}}$} \label{subsec: LIR}

The total integrated IR luminosity, $L_{\mathrm{IR}}$, measures the bolometric output of dust and has frequently been estimated from 24$\mu$m flux using MIR-IR conversion templates \citep[e.g.,][]{Wuyts2008,Whitaker2012,Whitaker2014}. At $z\sim1-3$, the Spitzer/MIPS 24$\mu$m band captures the rest-frame 7.7$\mu$m PAH feature \citep{Rieke2004}, which serves as a practical proxy for dust-obscured SFR via established PAH-to-dust-mass conversions \citep{Kennicutt2009, KennicuttEvans2012}. In our analysis, {\fontfamily{qcr}\selectfont Prospector} adopts the \citet{DraineLi2007} dust emission models to define the IR SED, whereas MIR-only conversions to $L_{\mathrm{IR}}$ typically rely on a single log-averaged template from \citet{DaleHelou2002} \citep[e.g.,][]{Whitaker2014}. To reduce template-dependent effects when analyzing the same underlying data set, \citet{Leja2022} fixed the \citet{DraineLi2007} templates to reproduce the $L_{\mathrm{8}}$ to $L_{\mathrm{IR}}$ color of the \citet{DaleHelou2002} templates, whereas we relax this constraint when incorporating FIR data. 

While the $L_{\mathrm{IR}}$/$L_{\mathrm{8}}$ ratio was found to remain approximately constant up to \textit{z} $\approx$ 2.5 \citep{Elbaz2011, Reddy2012}, \citet{Shivaei2024} recently used JWST/MIRI \citep{Rieke2015, Wright2023, Gardner2023} to show that the 7.7$\mu$m PAH strength correlates strongly with stellar mass. Although our $L_{\mathrm{IR}}$ and $L_{\mathrm{8}}$ quantities are derived from SED fits and are therefore model-dependent, they provide a direct bridge between our Herschel-constrained results and these JWST/MIRI measurements, motivating the comparison below.


We derive the total IR power ($F_{\mathrm{TIR}}$) by integrating the best-fit SED over the range 3-110$\mu$m ($\int F_{\lambda} \, d\lambda$). The integrated flux of the 7.7$\mu$m PAH feature ($F_{\mathrm{7.7}}$) is calculated following the methodology of \citet{Draine2021}. To isolate the PAH emission from the underlying warm dust continuum, we define a `clipped' flux by selecting anchor points on either side of the feature where the PAH emission begins. At the edge wavelengths (6.9 and 9.7$\mu$m for the 7.7$\mu$m feature), the flux is set to zero, and a `clip line' ($F_{\lambda}$) is drawn between them. The clipped flux for the 7.7$\mu$m feature is then defined as: 

\begin{equation}
F_{\text{clip}}(\text{band}) \equiv \int_{6.9\mu m}^{9.7\mu m} \left( F_{\lambda} - F_{\lambda}^{(\text{c.l.})} \right) \, d\lambda.
\end{equation}
\deleted{Visualizations of the regions used to calculate $F_{\mathrm{TIR}}$ and $F_{\mathrm{7.7}}$ are shown in Figure \ref{fig:mass stacked composites}.}

\begin{figure}[ht!]
\includegraphics[width=\columnwidth]{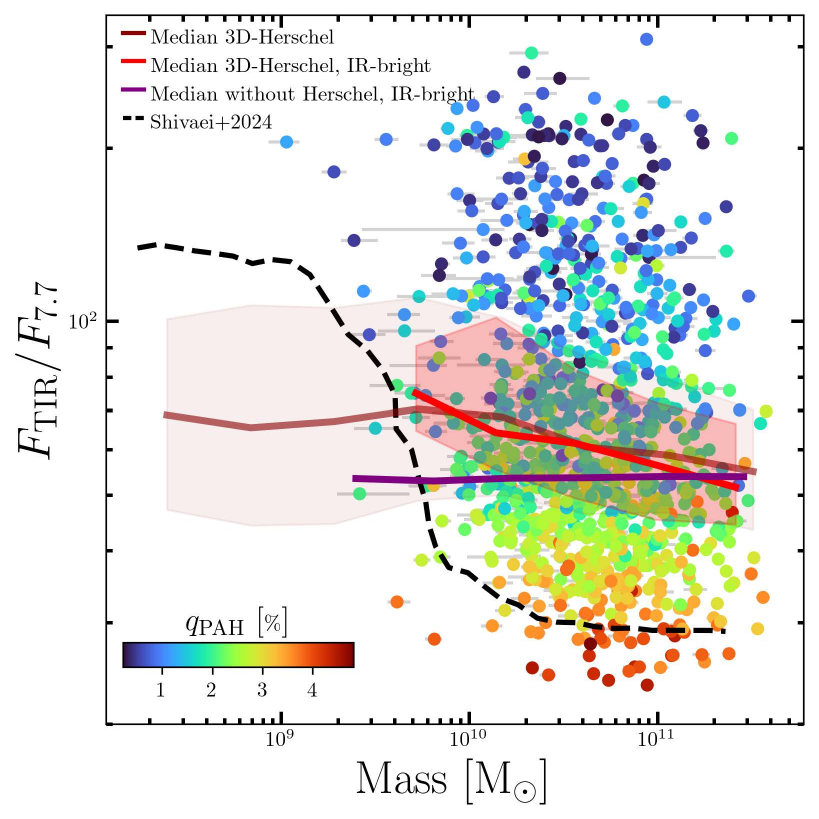}
\caption{
Estimates of $F_{\mathrm{TIR}}$/$F_{\mathrm{7.7}}$ as a function of stellar mass for the IR-bright sample, using Herschel constraints. Scatter points are color-coded by the fraction of dust mass in PAHs. Median relations are shown for the full 3D-Herschel sample (red) and for the IR-bright sample inferred without FIR data with a fixed template (purple; equivalent to the 24$\mu$m-to-70$\mu$m color of the log-average of the \citealt{DaleHelou2002} templates). The dashed line marks the mass-$F_{\mathrm{TIR}}$/$F_{\mathrm{7.7}}$ relation from \citet{Shivaei2024}.
}
\label{fig:FTIR-F77}
\end{figure}

We present our estimates of $F_{\mathrm{TIR}}$ and $F_{\mathrm{7.7}}$ for the IR-bright sample in Figure \ref{fig:FTIR-F77}, showing the ratio of the two as a function of stellar mass and color-coded by $q_{\mathrm{PAH}}$. The parameter $q_{\mathrm{PAH}}$ denotes the fraction of total dust mass in PAHs, serving as an analog to the
MIR-to-IR ratio. For context, we also plot the median relations for the IR-bright sample modeled without Herschel photometry and for the full 3D-Herschel star-forming population, with quiescent galaxies removed using a UVJ cut following \citet{Schreiber2015} and excluding galaxies with specific star formation rates (sSFR, SFR/$M_{*}$) $<10^{-10} \mathrm{yr^{-1}}$. This sample cut follows \citet{Shivaei2024}, allowing our study to be placed in context with theirs. 

The Herschel FIR constraints indicate that our previous fits, based on a fixed IR SED and a non-evolving $F_{\mathrm{TIR}}$/$F_{\mathrm{7.7}}$ ratio, underestimate the total IR emission at stellar masses $< 10^{10.4} M_{\odot}$. The relative strength of MIR emission decreases with stellar mass, indicating that diffuse dust increasingly dominates the energy budget over PAH re-radiation in obscured, low-mass galaxies. The $F_{\mathrm{TIR}}$/$F_{\mathrm{7.7}}$ ratio derived from 3D-Herschel fits displays substantial scatter, in contrast to the UV-MIR fits, which adopt fixed $F_{\mathrm{TIR}}$-to-$F_{\mathrm{7.7}}$ colors with a lower bound of $\sim$53 (median line, purple). At high stellar masses, the two methods converge, but in the lowest-mass bin (log($M_{\star}$) $\sim$ 9.6 $M_{\odot}$), the UV-MIR fits underestimate $F_{\mathrm{TIR}}$/$F_{\mathrm{7.7}}$ by $\sim$0.2 dex.

\deleted{However, we interpret this trend with caution: \citet{Elbaz2011} identified a tail of high $L_{\mathrm{IR}}$/$L_{\mathrm{8}}$ ratios associated with compact projected star formation densities, characteristic of galaxies in a starburst mode. The increased ``compactness'' of such systems produces more intense radiation fields, leading to PAH destruction and elevated $L_{\mathrm{IR}}$/$L_{\mathrm{8}}$ values. 

\citet{Elbaz2011} quantified starburst intensity using the excess of the specific SFR (i.e., the rate of growth of stellar mass per unit existing stellar mass, with sSFR$\equiv$SFR/M$_{\star}$) for a galaxy, labeled as ``starburstiness'' ($R_{\mathrm{SB}}$), defined by: 
\begin{equation}
    R_{\mathrm{SB}} = sSFR/sSFR_{\mathrm{MS}} 
\end{equation}
where $sSFR_{\mathrm{MS}}$ is the equivalent value for a galaxy of the same stellar mass residing on the SFMS for a given redshift. Galaxies are considered to be in a starburst mode when $R_{\mathrm{MS}} > 2$. To assess the role of starburst in our sample, we flag these objects and display them as stars in Figure \ref{fig:FTIR-F77}. Based on the Herschel-constrained fits, $\sim$6\% of our IR-bright sample qualifies as starbursts, slightly below the $\sim$7\% inferred without Herschel and considerably lower than the $<$20\% reported by \citet{Elbaz2011}.

In contrast to \citet{Elbaz2011}, we find that starburst galaxies do not elevate the median $F_{\mathrm{TIR}}$/$F_{\mathrm{7.7}}$ ratio within any stellar mass bin. Instead, they comprise the majority of our FIR-bright objects below log($M_{\star}/M_\odot) \sim 10$, reflecting both the limited depth of Herschel observations and the intrinsically FIR-bright nature of these systems. Including starburst galaxies is essential for extending the $F_{\mathrm{TIR}}$/$F_{\mathrm{7.7}}$ relation to lower stellar masses and for identifying where UV-MIR fits underestimate the IR luminosity.}

The offsets in $F_\mathrm{TIR}$/$F_\mathrm{7.7}$ are directly tied to how well the fixed $q_\mathrm{PAH}$ value reproduces the true PAH mass fraction. When the dust parameters are fixed in the UV-MIR fits such that $L_{8}$/$L_{\mathrm{IR}}$ follows the log-average of the \citet{DaleHelou2002} templates, fits that additionally include Herschel FIR photometry yield free $q_{\mathrm{PAH}}$ estimates that correlate strongly with offsets in the inferred IR luminosity. Specifically, 14$\%$ of IR-bright sources have $q_{\mathrm{PAH}}$ values significantly overestimated by the fixed prior, corresponding to an average increase in inferred $L_\mathrm{IR}$ of 0.27 dex and lower stellar mass estimates; conversely, 10$\%$ have $q_{\mathrm{PAH}}$ significantly underestimated, leading to an average decrease in $L_\mathrm{IR}$ of 0.14 dex.

The rise in $F_{\mathrm{TIR}}$/$F_{\mathrm{7.7}}$ toward lower stellar masses likely reflects the coupled influence of reduced PAH abundance in low-metallicity environments and PAH destruction by intense radiation fields \citep{Madden2006, Draine2007, Shivaei2017}, consistent with elevated $U_{\mathrm{min}}$ values we recover for nearly all IR-bright sources (see Section \ref{DustParams}).

\deleted{
To illustrate the increase in $F_{\mathrm{TIR}}$/$F_{\mathrm{7.7}}$ toward lower stellar masses, we present best-fit model composites of our IR-bright sample in bins of stellar mass in Figure \ref{fig:mass stacked composites}. Each composite SED is normalized at 1\micron\ and calculated following the same procedure as in Figures \ref{fig:1composite} and \ref{fig:composites}. For clarity, the regions under the SEDs corresponding to $F_{\mathrm{TIR}}$ and $F_{\mathrm{7.7}}$ are shaded, and the legend indicates the fractional contribution of each mass bin to the full 3D-Herschel mass-limited sample within that range.
}

Our $F_{\mathrm{TIR}}$/$F_{\mathrm{7.7}}$ estimates show reasonable agreement with \citet{Shivaei2024} at low stellar masses, but diverge by $\sim$0.28 dex at M$_\star \gtrsim 10^{10.3}$ M$_\odot$. To reconcile this, we leverage the overlap between the SMILES survey \citep{Rieke2024, Alberts2024} and our GOODS-South field, and directly compare inferred $L_{\mathrm{IR}}$ and $L_{\mathrm{7.7}}$ values: while $L_{\mathrm{IR}}$ shows good agreement (NMAD = 0.18 dex, negligible offset), $L_{\mathrm{7.7}}$ is systematically lower by 0.27 dex (NMAD = 0.27 dex). This offset is partially explained by model residuals of $\sim$0.1 dex in the MIPS 24$\mu$m fits, and by the fact that the \citet{Shivaei2024} analysis predates a significant MIRI flux calibration update. Differences in the $U_{\mathrm{min}}$ prior may also contribute: \citet{Shivaei2024} adopt an upper limit of 15, whereas we allow values up to 25, with $\sim$9\% of our IR-bright sample exceeding $U_{\mathrm{min}}$ = 15 and corresponding to elevated $F_{\mathrm{TIR}}$/$F_{\mathrm{7.7}}$ ratios.

\deleted{
\begin{figure*}
\plotone{composite_MassStacked.pdf}
\caption{ 
Composites from the IR-bright sample, binned by stellar mass, with the fractional contribution of each bin to the total sample for that mass range noted in the top left. Median $F_{\mathrm{TIR}}$/$F_{\mathrm{7.7}}$ values of each composite are listed on the right, showing a decreasing trend with stellar mass. Shaded regions indicate the areas of integration for $F_{\mathrm{TIR}}$ and $F_{\mathrm{7.7}}$.
\label{fig:mass stacked composites}}
\end{figure*}
}

\deleted{
\section{Discussion}  \label{sec:discussion}

This study represents one of the first large-scale applications of {\fontfamily{qcr}\selectfont Prospector} to incorporate FIR constraints from Herschel into UV–MIR SED fits for a mass-complete galaxy sample at $0.5 < z < 2.5$. By directly anchoring the dust emission with FIR data, we test the reliability of MIR-based estimates of obscured star formation and explore how FIR information alters inferences of stellar populations and dust properties. The primary limitation of this approach lies in the quality of the FIR photometry. Herschel imaging is hampered by its broad PSF and relatively shallow depth, leading to poor resolution and significant source blending (see Section \ref{subsec:ForcedPhotometry}). To mitigate these effects, we applied a deblending procedure to extract photometric measurements, improving FIR flux estimates for faint sources \citep{Lang2016}. Nevertheless, a substantial fraction of our measurements remain upper limits ($>$90\% in each Herschel band), particularly at the longest wavelengths. Despite these challenges, even upper limits provide meaningful constraints, as the addition of FIR photometry reduces uncertainties in key galaxy parameters. 

\subsection{Impact on the SFMS} \label{Impact On SFR-M Relation} 

Despite the challenges associated with Herschel data, we successfully incorporated FIR photometry (70-350$\mu$m) into the 3D-HST catalogs and fit the resulting 3D-Herschel sample using {\fontfamily{qcr}\selectfont Prospector}. The FIR observations provide direct measurements of dust emission, yielding more precise estimates of attenuation and related galaxy properties. Overall, Herschel data demonstrate that UV-MIR priors can reasonably predict FIR fluxes and recover key parameters across a broad population (Figure \ref{fig:cornerplot}). In our mass-limited sample of 34,443 galaxies, $\sim$70-80\% of objects have parameters consistent within 1$\sigma$ between the two sets of fits. 

Importantly, we find that stellar masses inferred from UV-MIR data alone are robust for $\sim80\%$ of galaxies, while SFRs are reliable for $\sim$70\%. As shown in Figure \ref{fig:IRbright_SFsequence}, the Herschel-constrained SFMS deviates by 0.1$\pm$0.07 dex on average relative to the UV-MIR inferred SFMS of \citet{Leja2022}, with no observed systematic offset. 
The latter SFMS is derived using a similar Bayesian inference with {\fontfamily{qcr}\selectfont Prospector}-$\alpha$ modeling (and effectively the same UV-MIR data). \citet{Leja2022} measure an SFMS with a normalization that is 0.2-0.5 dex lower than previous empirical estimates reported in the literature. Our results with FIR Herschel data generally corroborate these findings. 

Although we find greater IR-to-MIR luminosity ratios at low to intermediate stellar masses when incorporating Herschel FIR photometry (see Figure \ref{fig:FTIR-F77}), this increase in inferred SFRs appears mainly among the most highly obscured, IR-bright galaxies. For the majority of low-mass galaxies, the SFRs remain consistent with  previous UV-MIR estimates (Figure \ref{fig:IRbright_SFsequence}). 
Constraining robust SFRs at low stellar masses remains challenging, however, as the PAH features diminish in strength \citep{Shivaei2017}. The sensitivity and technical capabilities of JWST, enabling SFR measurements for a wide range of methodologies, will further illuminate this particular regime. Given the prevalence of upper limits in our sample, this study would also benefit from deeper, higher-quality FIR observations. While we may not have another space-based FIR observatory in our lifetime, this study demonstrates that the MIR can serve as a robust proxy for estimating key stellar parameters. 
}

\subsection{Dust emission parameters} \label{DustParams}

The dust emission parameters and dust temperatures derived in this section are tied to the \citet{DraineLi2007} framework. In particular, the temperatures we infer are characteristic dust temperatures from the best-fit models rather than effective dust temperatures derived from the FIR peak wavelength \citep[e.g.,][]{Elbaz2011, Casey2018, JonesStanway2023}. Absolute comparisons between these definitions are misleading, so our results should be interpreted in a relative sense — both internally, between our UV-MIR and UV-FIR fits, and externally against studies adopting the same parameterization. With these caveats in place, our constraints on $q_{\mathrm{PAH}}$ and $U_{\mathrm{min}}$ provide a benchmark for upcoming JWST/MIRI studies that will independently test the inferred PAH mass fractions and starlight intensities at the level of individual galaxies.

Incorporating FIR constraints enables a more informed analysis of the three dust emission parameters that shape the IR SED \citep{DraineLi2007}. In the absence of FIR data, it is common to fix these parameters or impose flat priors during modeling \citep{Leja2019b, Leja2022, Shivaei2024}, since MIR observations alone cannot fully constrain the total dust emission. The \citet{DraineLi2007} framework introduces $q_{\mathrm{PAH}}$, $U_{\mathrm{min}}$, and $\gamma$. Together, $U_{\mathrm{min}}$ and $\gamma$ determine the shape and peak wavelength of the FIR dust bump, with higher $U_{\mathrm{min}}$ corresponding to warmer dust temperatures and shorter peak wavelengths, and values near $U_{\mathrm{min}}\sim$1 indicate cooler dust temperatures \citep{Casey2018}. For a visualization of how these parameters influence the {\fontfamily{qcr}\selectfont Prospector}-modeled SED, see \citet{Leja2017}. 

\begin{figure}[ht!]
\includegraphics[width=\columnwidth]{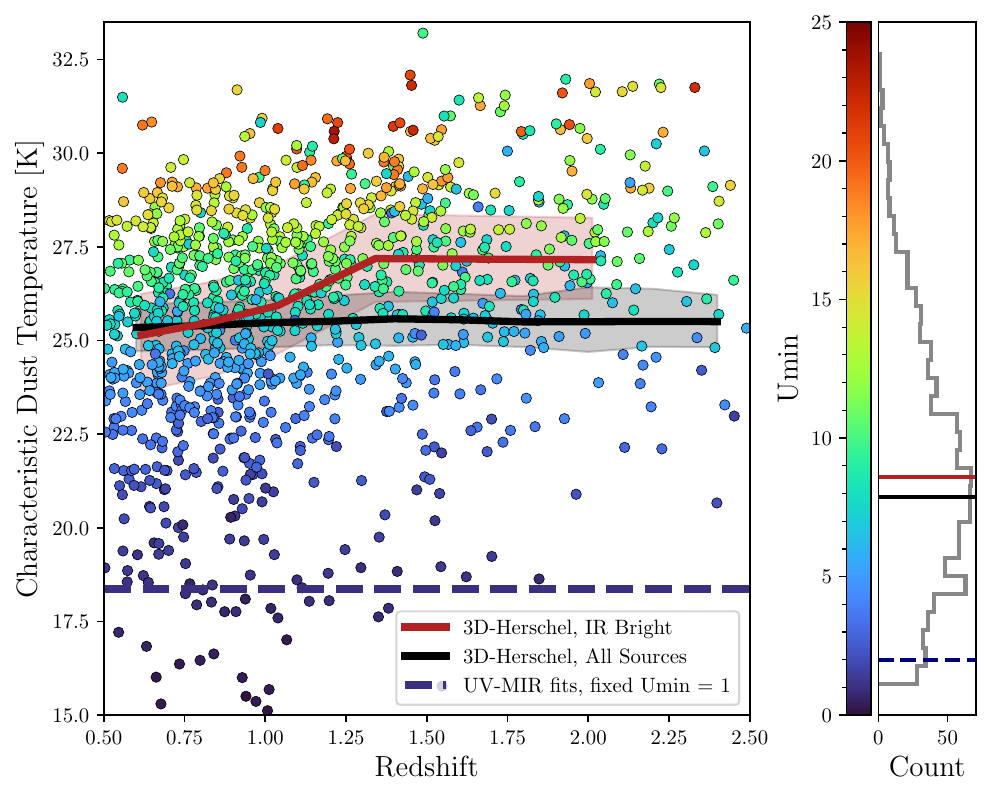}
\caption{
Characteristic dust temperature ($T_{d,char}$; Equation \ref{equation:Td}) as a function of redshift for the IR-bright sample. Points are color-coded by $U_{\mathrm{min}}$, with its distribution shown as a histogram along the color bar. Median reference lines mark the fixed $U_{\mathrm{min}}$ fits (dashed navy; 18.4K), the full 3D-Herschel mass-limited sample (black; $25.4 \ \pm \ 2.1$K), and the IR-bright subsample (red; $25.7 \ \pm \ 3.3$K).
  }
\label{fig:DustTempUmin}
\end{figure}

In our UV-MIR fits, the dust emission parameters are fixed to $U_{\mathrm{min}} = 1$, $q_{\mathrm{PAH}} = 2$, and $\mathrm{log(}\gamma) = 2$. Allowing them to vary in our 3D-Herschel fits shows that $q_{\mathrm{PAH}}$ and $\mathrm{log(}\gamma)$ remain consistent with their fixed values, with standard deviations of 0.16 and 0.45 dex, respectively. By contrast, $U_{\mathrm{min}}$ is systematically larger, with a mean value of $U_{\mathrm{min}} = 7.1 \pm 3.1$ -- about 0.85 dex higher than the fixed prior (right panel, Figure \ref{fig:DustTempUmin}). The fixed value $U_{\mathrm{min}}$$\sim$1 was originally motivated by preliminary tests in \citet{Leja2017} on 26 low-redshift KINGFISH disks \citep{Kennicutt2011}; we attribute the discrepancy with our higher value to that sample's restricted selection.

Dust temperature can be estimated first by calculating the mean starlight intensity $\langle U \rangle$, following Equation (33) of \citet{DraineLi2007}, which is dependent on $U_{\mathrm{min}}$, $U_{\mathrm{max}}$, and $\gamma$. Then we derive the characteristic dust temperature from Equation (13) of \citet{Draine2014}:

\begin{equation} \label{equation:Td}
    T_{\mathrm{d,char}} \approx 18 \ \langle U \rangle^{1/6} \ \mathrm{K}.
\end{equation}

For the IR-bright sample, we calculate an average luminosity-weighted dust temperature of $25.4 \ \pm \ 2.1\mathrm{K}$ (black line in Figure~\ref{fig:DustTempUmin}), systematically warmer than the fixed cold dust temperature of 18.4K adopted in the UV–MIR fits (dashed navy line). The scatter (red line) in Figure \ref{fig:DustTempUmin} illustrates the distribution of $U_{\mathrm{min}}$ values, with a noticeable step to warmer temperatures at $z\sim1-1.3$. This feature likely reflects a sampling bias, as Herschel preferentially detects higher-redshift star-forming galaxies with warm dust emission. The empirical composite SEDs (Figures \ref{fig:1composite}, \ref{fig:composites}), corroborate this discrepancy, with Herschel-informed spectra showing broader FIR dust bumps peaking $\sim$60$\mu$m shorter than the UV-MIR fits, indicating a broader underlying dust temperature distribution and systematically warmer characteristic dust temperatures.  


\deleted{
The discrepancy between the estimated dust temperatures is further reflected in our empirical composite SEDs with Herschel constraints (Figures \ref{fig:1composite}, \ref{fig:composites}). \deleted{, and \ref{fig:mass stacked composites}).}
In the left panel of Figure \ref{fig:composites}, we compare galaxies whose SFRs and dust attenuation remain consistent with previous UV-MIR fits: both composites show similar FIR emission strengths, but the Herschel-informed spectrum has a broader FIR bump that peaks $\sim$55\micron\ shorter. In the right panel, galaxies with enhanced dust emission and SFRs exhibit even broader bumps with peaks shifted by $\sim$62\micron\ relative to UV-MIR fits. These shifts towards shorter peak wavelengths indicate: (1) a broader underlying dust temperature distribution, and (2) a systematically warmer characteristic dust temperature \citep[see][]{Casey2012} compared to the model fits lacking FIR photometry.} 


The characteristic dust temperature of a galaxy depends on multiple factors -- including star formation activity, stellar population age, and the broader galaxy environment (e.g., the presence of AGNs \citealp{SomervilleDave2015}). Given the low resolution and shallow depth of Herschel imaging, this result requires validation with higher-resolution facilities such as ALMA or with next-generation FIR facilities such as PRIMA \citep{Carpenter2020, Moullet2023}.

\deleted{
\subsection{IR-bright sample}

We isolate a population of 1,118 IR-bright galaxies (Section \ref{sec:IR bright}), defined as sources with significant Spitzer/MIPS 24\micron\ detections (SNR$>$3) and detections in at least two Herschel bands. Compared to the full 3D-Herschel mass-limited samples, a larger fraction of these galaxies have best-fit parameters that fall outside the 1$\sigma$ limits ($>$57\% for offsets driven by dust properties; Figure \ref{fig:ADM degen}). Within the IR-bright sample, 73\% of stellar mass estimates and 66\% of SFR estimates remain consistent with the UV-MIR fits, slightly lower than the 78\% and 70\% agreement rates for the overall population of 34,443 galaxies. 

As shown in Figure \ref{fig:ADM degen}, the offsets are not unidirectional. Instead, a correlation between diffuse dust attenuation and SFR produces two comparably sized subpopulations: galaxies that are dustier, more star-forming, younger, somewhat less massive, and those that are less dusty, less star-forming, older, and more massive. Each subpopulation accounts for $\sim$8\% of the IR-bright sample. These results highlight that a substantial fraction of highly obscured systems (up to 21\% of our parent population; Table \ref{table:IRbright}) are poorly captured by fixed MIR-to-FIR templates, which fail to reproduce the diversity of IR colors in these galaxies.

We find that greater dust attenuation correlates with higher SFRs, consistent with theoretical expectations: stars enrich the ISM with dust both during their formation and at the end of their lifetimes, making these processes concurrent in actively star-forming galaxies. By contrast, dust is destroyed on relatively short timescales (on the order of 100 Myr) in quiescent systems \citep{Whitaker2021}, though additional dust can be produced more than a gigayear later during the AGB phase \citep{Conroy2013, hofner2018}. 

Incorporating dust emission measurements helps break the well-known degeneracy between dust, age, and (to a lesser extent at high redshift) metallicity. Reducing this degeneracy minimizes uncertainties in stellar mass estimates that otherwise arise when only UV-MIR data are available \citep{Conroy2013, Leja2019b}. Overall, our results demonstrate that {\fontfamily{qcr}\selectfont Prospector} UV-MIR fits provide a reliable tracer of the SFMS across a wide variety of galaxies, including most IR-bright objects when considering the average (Figure \ref{fig:IRbright_SFsequence}). At the lowest stellar masses within the IR-bright sample, however, there are hints of possible deviations. 

We identify a small number of low-mass galaxies (log($M_\star/M_{\odot}) <$ 9.8) at \textit{z} $\sim$ 1 with underestimated $\mathrm{SFR_{IR}}$ by the UV-MIR fits. \citet{Whitaker2017} show that the fraction of obscured star formation ($f_{\mathrm{obscured}}$) generally increases with stellar mass, but also reveal a tail of highly obscured low-mass galaxies at \textit{z} $<$ 1 and log($M_{\star}$/$M_{\odot}$) $\sim$ 9.4, where most galaxies exhibit $f_{\mathrm{obscured}}<$50\%. The obscured low-mass galaxies we identify contribute to this tail and have been largely overlooked in past studies, as their lack of PAH features \citep{Shivaei2017} produce systematic underestimates of FIR emission using rigid, MIR-dependent templates.
These extreme, low-mass starbursts comprise only a tiny fraction ($\lesssim$0.1\%) of the overall low-mass population, and MIR-to-IR conversions in the low-mass population as a whole remain model-dependent with large uncertainties \citep{Elbaz2011, Whitaker2017, Shivaei2020, Shivaei2024}.

The MIR range is dominated by PAH emission, with the strongest feature at 7.7$\mu$m. In our comparison fits without FIR constraints, the MIPS 24\micron\ band is used as a proxy for the total IR luminosity. For galaxies at 1.5 $<$ \textit{z} $<$ 2.5, this corresponds to rest-frame $\sim$8$\mu$m and is largely tracing the 7.7$\mu$m PAH feature. The reliability of using PAH emission as a tracer of total IR output has been the subject of ongoing debate \citep{Wuyts2008, Elbaz2011, Reddy2012, Shivaei2017}. Recent work has shown that the ratio $F_{\mathrm{TIR}}$/$F_{\mathrm{7.7}}$ correlates with stellar mass \citep{Shivaei2024}, likely reflecting the dependence of PAH abundance on stellar metallicity \citep{Madden2006, Hunt2010, Seok2014, XieHo2019, Li2020} together with the well-established mass-metallicity relation \citep{Tremonti2004, Topping2021}. 

The formation and destruction of PAHs in the ISM is complex and involves multiple pathways. In the following subsection, we explore how these processes shape the observed offsets in $F_{\mathrm{TIR}}$/$F_{\mathrm{7.7}}$ and place our results in the context of \citet{Shivaei2024}. 
}

\deleted{
\subsection{MIR-to-IR luminosity relationship}

In our analysis, the models constrained only by UV-MIR photometry convert Spitzer/MIPS 24$\mu$m fluxes into total IR luminosity using the models from \citet{DraineLi2007} with parameters fixed such that $L_{8}/L_{\mathrm{IR}}$ follows the log-average of the \citet{DaleHelou2002} templates. At redshifts $z \sim 1.5$–2.5, the Spitzer/MIPS 24$\mu$m bandpass captures the strong 7.7$\mu$m PAH feature, which contributes $\sim$40-50\% of the total PAH emission \citep{Hunt2010} and serves as a widely used proxy for MIR output. However, studies have shown that relying on MIR emission to approximate $L_{\mathrm{IR}}$ and derive $\mathrm{SFR_{IR}}$ \citep[e.g.,][]{Daddi2007, Wuyts2008, Whitaker2014}, systematically underestimates both by a factor of $\sim$2 for galaxies with $M_*$ $\lesssim$ $10^{10}$$M_{\odot}$ and fails to recover $L_{\mathrm{IR}}$ accurately beyond \textit{z} $\sim$ 2 \citep{Elbaz2011, Schreiber2015, Shivaei2017}. By contrast, we find that including FIR photometry in our fits results in accurate $\mathrm{SFR_{IR}}$ and $L_{\mathrm{IR}}$ inferences for a majority of our parent sample.

As shown in Figure \ref{fig:FTIR-F77}, the largest offsets in the $F_{\mathrm{TIR}}$/$F_{\mathrm{7.7}}$ ratio -- equivalent to $L_{\mathrm{IR}}$/$L_{\mathrm{8}}$ -- occur around log($M_{\star}$/$M_{\odot}$) $\sim$ 9.6, near the lower bound of our IR-bright sample. In this regime, Herschel-constrained fits yield $F_{\mathrm{TIR}}$/$F_{\mathrm{7.7}}$ ratios larger by $\sim$0.2 dex compared to the UV-MIR fits, suggesting that {\fontfamily{qcr}\selectfont Prospector} models with fixed MIR-IR colors may systematically underpredict the total IR strength in low-mass systems.

The fixed MIR-IR colors in our models are governed by the three dust emission parameters---$\gamma$, $U_{\mathrm{min}}$, and most importantly, $q_{\mathrm{PAH}}$, which represents the fraction of total dust mass in PAHs and is effectively equivalent to the ratio of MIR to total IR luminosity. When the dust parameters are fixed in the UV-MIR fits such that $L_{8}/L_{\mathrm{IR}}$ follows the log-average of the \citet{DaleHelou2002} templates, fits that additionally include Herschel FIR photometry yield free $q_{\mathrm{PAH}}$ estimates that exhibit a strong correlation with offsets in the inferred IR luminosity. Specifically, 14\% of IR-bright sources show $q_{\mathrm{PAH}}$ values that are significantly overestimated when adopting a previously fixed value of $q_{\mathrm{PAH}}=2$ (offsets exceeding the $1\sigma$ uncertainty relative to the fixed-$q_{\mathrm{PAH}}$ fits), corresponding to an average increase in the inferred IR luminosity of $0.27$ dex and lower stellar mass estimates. Conversely, 10\% of the IR-bright sample have $q_{\mathrm{PAH}}$ values that are significantly underestimated in the UV-MIR fits, leading to an average decrease in $L_{\mathrm{IR}}$ of $0.14$ dex.

Previous work has found that the $L_{\mathrm{IR}}$/$L_{\mathrm{8}}$ ratio remains nearly constant over most of cosmic history. Using the IRAC 8 $\mu$m bandpass, \citet{Elbaz2011} measured $L_{\mathrm{IR}}/L_{8} = 4.9^{+2.9}{-2.2}$ out to $z\sim2.5$, while \citet{Reddy2012} reported $7.7 \pm 1.6$ over a similar redshift range. A nearly constant $L_{\mathrm{IR}}$/$L_{\mathrm{8}}$ ratio implies little evolution in the IR SED over this redshift range, a crucial factor when interpreting galaxy SEDs with limited IR coverage. Our results suggest, however, that 
low-mass IR-bright galaxies may deviate from this picture, highlighting the importance of direct FIR constraints in this regime.

The link between MIR emission and total IR output is not universal, as it depends critically on the abundance of PAHs. FIR templates are often used under the assumption that MIR emission scales proportionally with total IR emission across all galaxy types, but PAH production and survival are regulated by a galaxy's physical and environmental conditions. PAHs are produced primarily in the ejecta of supernovae and in the carbon-rich outflows of AGB stars \citep{FrenklachFeigelson1989, Galliano2008, hofner2018}. Their formation is more efficient in high-metallicity environments, where shattering of carbonaceous dust grains replenishes the PAH population \citep{Seok2014}. On the other hand, low-metallicity systems exhibit reduced PAH production: either because young chemical environments lack sufficient gas-phase carbon to form PAHs \citep{Draine2007}, or because PAHs are destroyed by thermal sputtering in shock-heated, metal-poor gas with inefficient cooling \citep{Li2020}. In addition, in environments without sufficient dust shielding, intense radiation fields can rapidly destroy PAHs \citep{Madden2006, Hunt2010, Magdis2013, Shivaei2017, XieHo2019}.

In our modeling of the IR-bright sample, Nearly all sources exhibit elevated $U_{\mathrm{min}}$ values, which also implies a suppression of $F_{\mathrm{7.7}}$ relative to $F_{\mathrm{TIR}}$, consistent with expectations that PAHs are destroyed in stronger radiation fields \citep{Madden2006, Hunt2010, Magdis2013, Shivaei2017, XieHo2019}. We also find tentative evidence that some lower-metallicity galaxies exhibit increased SFRs (Figures \ref{fig:cornerplot} and \ref{fig:ADM degen}). These systems show higher $F_{\mathrm{TIR}}$-to-$F_{\mathrm{7.7}}$ ratios, consistent with the idea that metals provide dust shielding against intense UV radiation \citep{Hunt2010, Magdis2013, Shivaei2017}. Together, these trends suggest an interconnected picture: enhanced IR emission boosts the inferred SFRs, while stronger radiation fields in low-metallicity environments both heat the dust and suppress PAH survival. The observed rise in $F_{\mathrm{TIR}}$/$F_{\mathrm{7.7}}$ therefore reflects the coupled influence of dust heating, metallicity, and PAH destruction.
}

\deleted{
\subsection{Systematic Differences in $F_{\mathrm{TIR}}/F_{7.7}$}

In Figure \ref{fig:FTIR-F77}, the $F_{\mathrm{TIR}}$/$F_{\mathrm{7.7}}$ ratios estimated from fits with and without Herschel photometry agree at the massive end. However, the median trend from \citet{Shivaei2024} asymptotes to value that is lower by $\sim$0.28 dex relative to our measurements for galaxies $\gtrsim$10$^{10.3}$ M$_{\odot}$. Their analysis, based on JWST/MIRI data, compares the 7.7$\mu$m PAH flux to the total IR flux (3-1100$\mu$m) of star-forming galaxies using eight JWST/MIRI broadband filters spanning 5.6-25.5$\mu$m. Our study, by contrast, relies on a single Spitzer/MIPS 24\micron\ measurement, which traces the rest-frame 7.7$\mu$m feature only for galaxies at 1.5 $\lesssim$ \textit{z} $\lesssim$ 2.5. \citet{Shivaei2024} further find that the 7.7$\mu$m PAH strength correlates strongly with stellar mass, following a logistic relation in which PAH emission increases towards higher masses, whereas we do not recover a strong stellar mass dependence in our work. 

To investigate the origin of this discrepancy, we directly compare the inferred values from our work to that in \citet{Shivaei2024}.  Fortuitously, the Systematic Mid-infrared Instrument Legacy Extragalactic Survey \citep[SMILES;][]{Rieke2024, Alberts2024} analyzed in \citet{Shivaei2024} lies within the GOODS-South field, which is also covered by 3D-Herschel, enabling a direct comparison between our inferred $L_{\mathrm{IR}}$ and $L_{\mathrm{7.7}}$ values and those reported by \citet{Shivaei2024}. We quantify the scatter in both $L_{\mathrm{IR}}$ and $L_{\mathrm{7.7}}$ using the normalized median absolute deviation (NMAD). While the $L_{\mathrm{IR}}$ estimates show an NMAD of 0.18 dex with negligible median offset, the $L_{\mathrm{7.7}}$ values exhibit a larger scatter (NMAD = 0.27 dex) and are 0.27 dex systematically lower than those reported by \citet{Shivaei2024}.

To test for calibration issues within our data, We examine the residuals between the observed MIPS 24$\mu$m fluxes and the best-fit model prediction at the same wavelength. We find that the model fluxes are approximately 0.1 dex lower than the observed MIPS 24$\mu$m fluxes, consistent with the lower $L_\mathrm{7.7}$ values we estimate relative to those in \citet{Shivaei2024}. However, while this alleviates some tension, this does not fully account for the discrepancy. It is also worth noting that the \citet{Shivaei2024} analysis predates a significant update to the MIRI calibration, which resolved a systematic flux offset and may contribute to their higher inferred PAH luminosities. 

Differences in the treatment of the dust-heating parameter $U_{\mathrm{min}}$ may further contribute to the observed offset. In particular, \citet{Shivaei2024} adopt a flat prior for $U_{\mathrm{min}}$ between 0.1-15 \citep{Leja2017}, whereas our modeling imposes a more conservative upper limit of 25. Roughly 9\% of our IR-bright sample has $U_{\mathrm{min}} > 15$, corresponding to higher $F_{\mathrm{TIR}}$/$F_{\mathrm{7.7}}$ values. While tighter priors on dust emission parameters can reduce biases and uncertainties in IR estimates without FIR data \citep{Leja2017}, as is the case for the \citet{Shivaei2024} analysis, allowing $U_\mathrm{min}$ to exceed 15 provides flexibility to capture galaxies with more intense radiation field and elevated $F_{\mathrm{TIR}}$/$F_{\mathrm{7.7}}$ ratios.

Taken together, the tension between this $F_{\mathrm{TIR}}/F_{\mathrm{7.7}}$ ratios at intermediate-to-high stellar masses likely results from some combination of the following: (1) the $U_{\mathrm{min}}$ prior adopted in \citet{Shivaei2024} may bias the average $F_{\mathrm{TIR}}/F_{\mathrm{7.7}}$ ratios low, and (2) there may be residual systematics in the MIR flux calibrations in both studies. While we cannot definitively isolate the origin of this offset in $F_{\mathrm{TIR}}/F_{\mathrm{7.7}}$, this comparison highlights the importance of combining high-quality JWST data with wide-field, multi-survey analyses like ours to fully reconcile these measurements. Future facilities such as PRIMA will be critical for reopening a panoramic view of the MIR-to-FIR universe and for resolving remaining systematics in PAH-based IR diagnostics.
}

\section{Conclusions}  \label{sec:conclusion}

In this work, we incorporated FIR Herschel photometry (70 - 350\micron) into SED fitting of the 3D-HST photometric catalogs using the Bayesian modeling code {\fontfamily{qcr}\selectfont Prospector}. Because of the shallow depth and large PSF of the Herschel images, flux extraction requires sophisticated deblending techniques. Here, we perform `forced photometry' with {\fontfamily{qcr}\selectfont MOPHONGO} and {\fontfamily{qcr}\selectfont T-PHOT}, enabling recovery of FIR fluxes down to fainter limits than otherwise possible. These measurements were added to the 3D-HST photometric catalog to produce a comprehensive, 0.3 - 350$\mu$m dataset, which we term 3D-Herschel and release publicly with this paper.

We validated the 3D-Herschel fluxes and source counts against independent Herschel analyses (\textsection \ref{sec:validation}). Flux comparisons (Figure \ref{fig:flux comparisons}) show consistency with published catalogs, while galaxy counts across the four fields and Herschel bands (Figure \ref{fig:galaxy counts}) confirm the reliability of the deblending pipeline.

With the validated 3D-Herschel catalogs, we performed SED fitting spanning the UV-FIR using {\fontfamily{qcr}\selectfont Prospector}. The addition of FIR data provided direct constraints on dust emission, allowing us to expand to a 17-parameter model that includes the three dust emission parameters -- $q_{\mathrm{PAH}}$, $\gamma$, and $U_{\mathrm{min}}$ -- which were previously fixed when modeling relied only on UV-MIR data.

The main findings of our analysis are summarized as follows:
\begin{enumerate}

    \item The addition of Herschel FIR constraints to the fiducial UV-MIR fits does not significantly alter the majority of sources. Fitting UV-MIR photometry alone with {\fontfamily{qcr}\selectfont Prospector} reliably infers key galaxy properties and predicts $L_{\mathrm{IR}}$, with $\sim$two-thirds of parameter offsets between UV-MIR and UV-FIR fits falling within the 1$\sigma$ uncertainties. 
    
    \item The SFMS inferred from UV-FIR photometry is consistent with that derived from UV-MIR-only {\fontfamily{qcr}\selectfont Prospector-$\beta$} fits, with an average deviation of 0.1$\pm$0.07 dex at fixed stellar mass and no more than 0.25 dex in any individual mass bin.  This agreement confirms that the $\sim$0.2-0.5 dex offset between the {\fontfamily{qcr}\selectfont Prospector-$\beta$} SFMS and traditional UV+IR estimates reported by \citet{Leja2022} is robust to the lack of direct FIR constraints.
    
    \item Incorporating FIR photometry enables three additional free parameters -- $q_{\mathrm{PAH}}$, $\gamma$, and $U_{\mathrm{min}}$ -- to be constrained, whereas in the UV-MIR models they were fixed to correspond to cold dust temperatures, while our FIR-constrained field yield an average $U_{\mathrm{min}}=7.1$, implying dust temperatures $\sim$7K warmer.
    
    \item UV-MIR models rely on fixed IR templates that assume a constant $F_{\mathrm{TIR}}/F_{\mathrm{7.7}}$ ratio across stellar mass. By contrast, fits to the IR-bright sample (spanning 0.1\% of mass-limited sample at log($M_{\star}/M_{\odot}$) = 9.5 to $\sim$20\% for log($M_{\star}/M_{\odot}) \geq$ 11.25) show that $F_{\mathrm{TIR}}$/$F_{\mathrm{7.7}}$ increases toward lower stellar masses, with a $\sim$0.2 dex offset from the UV-MIR fits at log($M_{\star}$/$M_{\odot}$) $\sim$ 9.6. This demonstrates that fixed IR templates underpredict FIR emission in the most obscured, low- to intermediate-mass galaxies.

\end{enumerate}

The Herschel Space Observatory represents a past generation of space-based telescopes that opened a crucial window into the IR universe, albeit at limited resolution. In the upcoming era of advanced IR facilities, galaxy evolution studies will benefit from more sensitive, higher-resolution data, particularly in the MIR–FIR regime that remains largely inaccessible today. Our results, derived primarily from upper-limit constraints, highlight the substantial gains to be made in measuring key galaxy scaling relations with improved data. Looking ahead, future space missions with MIR–FIR coverage, together with FIR–to–submillimeter observations from facilities such as ALMA, will be essential to fully characterize the dust and star formation properties of galaxies.

\vspace{0.5em}
\noindent\centerline{\textbf{Data Availability}}
\vspace{0.025em}

The UV-to-FIR photometric catalogs presented in this work are publicly available on Zenodo at \dataset[doi:10.5281/zenodo.20706940]{https://doi.org/10.5281/zenodo.20706940}.

\vspace{0.5em}
\noindent\centerline{\textbf{Acknowledgments}}
\vspace{0.025em}

We thank the anonymous referee for their constructive comments, which improved the quality of this manuscript. This work is based on observations taken by the 3D-HST Treasury Program (GO 12177 and 12328) with the NASA/ESA HST, which is operated by the Association of Universities for Research in Astronomy, Inc., under NASA contract NAS5-26555. This work makes use of the 3D-HST photometric and grism spectroscopic catalogs \citep{Skelton2014, Momcheva2016, Brammer2012a}, which are publicly available from the Mikulski Archive for Space Telescopes (MAST) at https://archive.stsci.edu/prepds/3d-hst/. This work is based in part on observations made with the Spitzer Space Telescope, which was operated by the Jet Propulsion Laboratory, California Institute of Technology, under a contract with NASA. Herschel is an ESA space observatory with science instruments provided by European-led Principal Investigator consortia and with important participation from NASA. The Herschel data used in this work were obtained as part of the GOODS-Herschel (PI: D. Elbaz), CANDELS-Herschel (PI: M. Dickinson), PACS Evolutionary Probe (PEP), and HerMES programs.

Support for this work was provided by NASA through the Astrophysics Data Analysis Program (ADAP) grant 80NSSC20K0416. Our research is made possible by the use of the panchromatic SED modeling code {\fontfamily{qcr}\selectfont Prospector}. Additionally, the use of the FSPS code in {\fontfamily{qcr}\selectfont Prospector} has allowed us to compare robust measurements of physical galaxy properties. We would like to thank Stacey Alberts for their support in this study. We acknowledge the EAZY SED modeling code as well which our photometric redshifts have been taken from.

This research made use of astrodendro, a Python package to compute dendrograms of Astronomical data \href{http://www.dendrograms.org/}{http://www.dendrograms.org/}, as well as Astropy, a community-developed core Python package for Astronomy (Astropy Collaboration, 2013), \href{http://www.astropy.org}{http://www.astropy.org}.

\textit{Software}: Prospector \citep{JohnsonLeja2017}, pythonfsps \citep{Foreman-Mackey2014}, Astropy \citep{Astropy2013, Astropy2018}, FSPS \citep{Conroy2009},
matplotlib \citep{Caswell2020}, scipy \citep{Pauli2020}, ipython \citep{PerezGranger2007}, numpy \citep{VanDerWalt2011}.


\appendix
\section{Flux Validation and Photometric Consistency}
\label{app:validation}
\restartappendixnumbering

{\normalsize 
\begin{deluxetable*}{lcccc}
\label{table:3D-Herschel}
\tablecaption{3D-Herschel Comparison Samples to Validate the Photometry}
\tablewidth{0pt}
\tablehead{
\colhead{} &
\colhead{GOODS-S} &
\colhead{GOODS-N} &
\colhead{COSMOS} &
\colhead{UDS}
}
\startdata
Total Number of Objects
& 50,507 & 38,279 & 33,879 & 44,102 \\
\hline
\addlinespace[3pt]
PACS Comparison Catalog
& \citet{Magnelli2013}
& \shortstack[c]{GOODS-N\\``Super-deblended" catalog}
& \shortstack[c]{COSMOS\\``Super-deblended" catalog}
& \citet{Oliver2012} \\
\addlinespace[0.25pt]
SPIRE Comparison Catalog
& \citet{Shirley2021}
& \citep{Liu2018}
& \citep{Jin2018}
& \citet{Oliver2012} \\
\enddata
\tablecomments{For GOODS-North and COSMOS, the same super-deblended catalogs were used for both PACS and SPIRE comparisons.}
\end{deluxetable*}
}

We validate our FIR photometry against a heterogeneous set of public Herschel catalogs that span a range of extraction methodologies and imaging depths.  
In GOODS-South, we compare against the PACS catalogs of \citet{Magnelli2013}, which combine PEP and GOODS-Herschel observations to reach 3$\sigma$ depths of $\sim$0.6-1.3 mJy at 100 and 160$\mu$m in the deepest regions, among the deepest PACS blank-field data available, and against \added{the prior-based XID+ SPIRE extractions of \citet{Shirley2021}. In UDS, comparison fluxes are drawn from the publicly released HerMES catalog products\footnote{\texttt{UDS\_PACSxID24\_v1} and \texttt{L4-UDS\_xID250\_DR2}} \citep{Oliver2012}. In GOODS-North and COSMOS, we compare against the super-deblended catalogs of \citet{Liu2018} and \citet{Jin2018}, which apply prior-driven source extraction to deep Herschel imaging that substantially overlaps with the datasets analyzed here. In GOODS-North, both PACS and SPIRE photometry are derived from GOODS-Herschel, with PACS supplemented by PEP. In COSMOS, the PACS imaging combines CANDELS-Herschel and PEP, while the SPIRE measurements are drawn from HerMES. Taken together, these catalogs span both prior-driven super-deblending and more traditional extraction approaches, providing a stringent test of our photometry across all four fields.}

\subsection{Mean Flux Offsets}

\begin{figure*}[ht!]
\centering
\includegraphics[width=\textwidth]{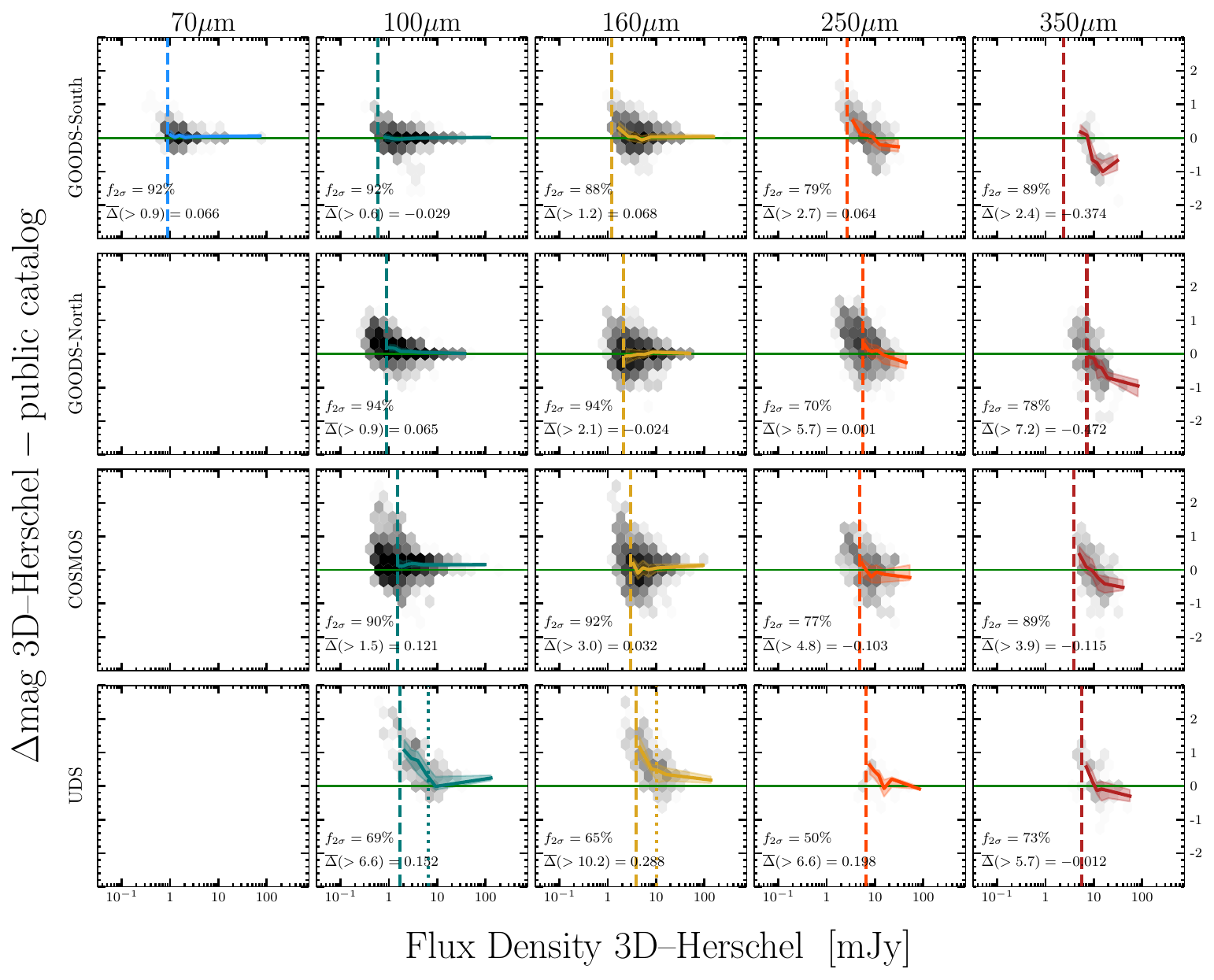}
\caption{Overall agreement of 3D-Herschel fluxes compared to published Herschel catalogs. The background color scale shows the logarithmic number density of sources in our sample. The dashed line indicates the median 3$\sigma$ flux uncertainty from the original survey publications, whereas dotted lines indicate limits we infer herein for sources originally detected with Spitzer, as opposed to the forced photometry methods assumed in this study (Table \ref{tab:herscheldata}). The mean offset for objects above the 3$\sigma$ flux uncertainty (dashed line, or dotted if present) is listed in the lower left of each panel.
\label{fig:flux comparisons}}
\end{figure*}

We match sources to the nearest object within a fixed radius and compare fluxes (Figure \ref{fig:flux comparisons}). We apply an SNR $>$ 1 cut to both the comparison catalog and the 3D-Herschel catalog to retain sources with marginal individual-band detections, as appropriate for confusion-limited FIR data (roughly 23\% of the 3D-Herschel parent sample). For COSMOS, GOODS-N, and GOODS-S, we adopt a 0.5 arcsecond radius since the parent catalogs are based on high-resolution imaging priors. For UDS, we use a 3 arcsecond radius, as sources are detected only at FIR wavelengths where positional uncertainties are comparable to the PACS beam size.

In general, the 3D-Herschel catalog is in good agreement with published catalogs, indicating the deblending process was successful and fluxes are robustly extracted. We find mean offsets of $\lesssim$0.3 mag at 100-250\micron\, with most fields/filters agreeing within $<$0.15 mag. The 350\micron\ 3D-Herschel photometry shows notably larger offsets (ranging from $\sim$0.03 mag in some fields to as much as 1 mag in others), likely due to the shallow depth of this band and differences in flux extraction methods between catalogs. Despite these offsets, the boosted 350$\mu$m uncertainties typically exceed these systematic offset, so the SED fits weight this band accordingly.

\subsection{Individual Source Consistency}

\begin{figure}[ht!]
\centering
\includegraphics[width=\columnwidth,height=0.7\textheight,keepaspectratio]{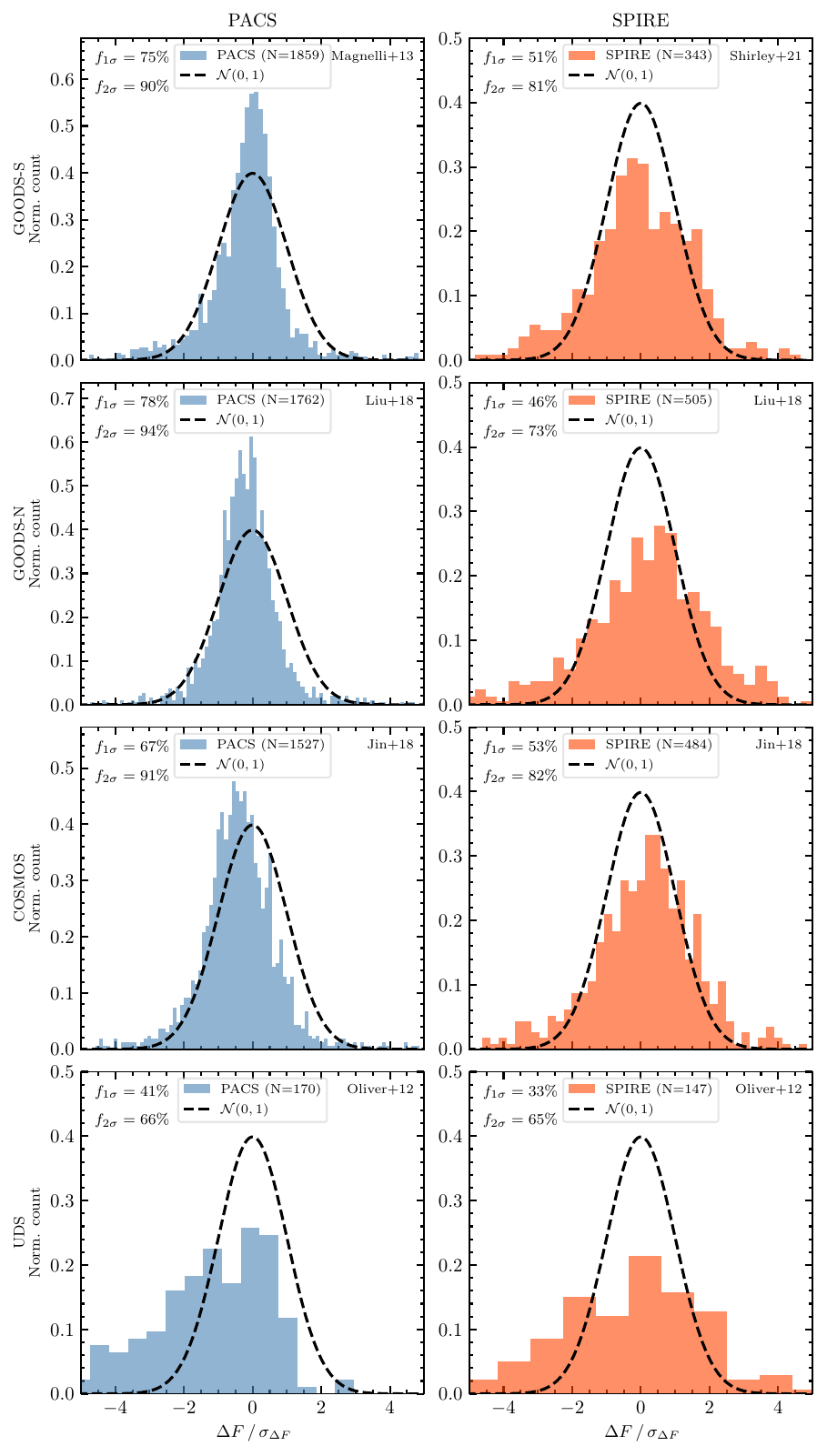}
\caption{Distributions of uncertainty-normalized flux residuals 
(Equation~\ref{eq:norm_resid_flux}) for each field and instrument. Left 
column: PACS bands; right column: SPIRE bands. Rows correspond to 
GOODS-S, GOODS-N, COSMOS, and UDS from top to bottom. The dashed curve 
shows the unit Gaussian $\mathcal{N}(0,1)$ expected for perfectly 
Gaussian, well-calibrated uncertainties. The comparison catalog used in 
each panel is indicated in the upper right. The fractions $f_{1\sigma}$ 
and $f_{2\sigma}$ give the fraction of sources with 
$|\Delta F / \sigma_{\Delta F}|$ within 1 and 2, respectively.
\label{fig:norm_resid}}
\end{figure}

While the flux ratios above test for population-level systematics, they do not assess whether per-source uncertainties are correctly calibrated. To do this, we compute the uncertainty-normalized flux residual for each matched source,
\begin{equation}
    \frac{\Delta F}{\sigma_{\Delta F}} = 
    \frac{F_\mathrm{3DH} - F_\mathrm{comp}}{\sqrt{\sigma^2_\mathrm{3DH} 
    + \sigma^2_\mathrm{comp}}}
    \label{eq:norm_resid_flux}
\end{equation}
where $F_\mathrm{3DH}$ and $F_\mathrm{comp}$ are the 3D-Herschel and 
comparison catalog flux densities respectively, and $\sigma_\mathrm{3DH}$, 
$\sigma_\mathrm{comp}$ are their associated uncertainties. This is 
analogous to Equation \ref{eq:flux_resid_param}, adapted here for flux densities rather than 
derived physical quantities. For a sample with well-characterized Gaussian 
uncertainties, one would expect $\sim$68\% and $\sim$95\% of sources to 
fall within $|\Delta F / \sigma_{\Delta F}| < 1$ and $< 2$ respectively.

Figure \ref{fig:norm_resid} shows the distribution of 
$\Delta F / \sigma_{\Delta F}$ for each instrument and field combination, 
with each panel combining all available bands for that instrument (PACS: 
100 and 160\micron, or 70, 100 and 160\micron\ for GOODS-S; SPIRE: 250 
and 350\micron). The overplotted $\mathcal{N}(0,1)$ Gaussian provides a 
reference for ideally calibrated errors.

The residual distributions exhibit two features that together explain the 
$f_{2\sigma}$ values falling below the 95.4\% Gaussian expectation. First, 
the central peaks are narrow and tall relative to $\mathcal{N}(0,1)$, 
indicating that the majority of matched sources have well-behaved, 
consistent flux measurements between catalogs. Second, non-Gaussian 
shoulders and tails are present in all panels, reflecting a minority 
population of sources with larger-than-expected residuals that pull the 
enclosed fraction below the Gaussian expectation. For the PACS 
distributions in GOODS-S, GOODS-N, and COSMOS, the widths of the 
distributions are modestly broader than the expected Gaussian, and a mild negative offset is present, suggesting that our fluxes are on average slightly 
lower than the comparison catalogs -- consistent with prior-based 
deblending methods recovering less flux than methods that do not 
simultaneously fit neighboring sources. The SPIRE distributions are 
broader still, with more prominent tails, which is expected given the 
confusion-limited nature of SPIRE imaging and the inherent difficulty of 
characterizing per-source uncertainties in the prior-extraction regime.

The UDS field shows broader residuals and a systematic negative offset in both PACS ($f_{2\sigma} \approx 66\%$) and SPIRE ($f_{2\sigma} \approx 65\%$). Our UDS catalog relies on FIR-based positional estimates, necessitating a larger 3 arcsecond matching radius to account for positional uncertainties comparable to that of FIR-detected sources. At this scale, chance associations are more common, where a brighter neighboring source at the edge of the matching radius may be selected, biasing the matched comparison fluxes systematically high relative to our measurements and producing the observed negative offset. The broader tails reflect genuine source confusion at this angular scale, where multiple sources contribute flux within the beam. UDS sources are retained in the analysis, but results that depend sensitively on individual UDS flux measurements should be interpreted with this caveat in mind.


\bibliography{bib}{}
\bibliographystyle{aasjournal}

\listofchanges

\end{document}